%% file: main.tex
\documentclass[conference]{IEEEtran}
\IEEEoverridecommandlockouts
\usepackage[T1]{fontenc}
\usepackage[utf8]{inputenc}
\usepackage[final,nopatch=footnote]{microtype}
\usepackage{cite}
\usepackage{array}
\usepackage{amsmath,amssymb,amsfonts}
\usepackage{algorithmic}
\usepackage{cite}
\usepackage{graphicx}
\usepackage{textcomp}
\usepackage{xcolor}
\usepackage{xspace}
\usepackage{hyperref}
\usepackage{cleveref}
\usepackage{setspace}
\usepackage{mathtools}
\usepackage{tcolorbox}
\usepackage{multicol}
\usepackage[font=footnotesize]{caption}
\usepackage{balance}

\balance

\DeclarePairedDelimiter\ceil{\lceil}{\rceil}

\def\BibTeX{{\rm B\kern-.05em{\sc i\kern-.025em b}\kern-.08em
    T\kern-.1667em\lower.7ex\hbox{E}\kern-.125emX}}
\begin{document}

\input{misc/macros}

\title{On Accelerating Deep Neural Network Mutation Analysis by Neuron and Mutant Clustering}

\author{
\IEEEauthorblockN{Lauren Lyons}
\IEEEauthorblockA{
\textit{Auburn University}\\
Auburn, AL, USA \\
lauren.n.lyons@auburn.edu}
\and
\IEEEauthorblockN{Ali Ghanbari}
\IEEEauthorblockA{
\textit{Auburn University}\\
Auburn, AL, USA \\
ghanbari@auburn.edu}
}

\maketitle

\begin{abstract}

Mutation analysis of deep neural networks (DNNs) is a promising method for effective evaluation of test data quality and model robustness, but it can be computationally expensive, especially for large models.
To alleviate this, we present \dmaacc, a technique and a tool that speeds up DNN mutation analysis through neuron and mutant clustering.
\dmaacc implements two methods: (1) neuron clustering to reduce the number of generated mutants and (2) mutant clustering to reduce the number of mutants to be tested by selecting representative mutants for testing.
Both use hierarchical agglomerative clustering to group neurons and mutants with similar weights, with the goal of improving efficiency while maintaining mutation score.

\dmaacc has been evaluated on 8 DNN models across 4 popular classification datasets and two DNN architectures.
When compared to exhaustive, or vanilla, mutation analysis, the results provide empirical evidence that neuron clustering approach, on average, accelerates mutation analysis by 69.77\%, with an average -26.84\% error in mutation score.
Meanwhile, mutant clustering approach, on average, accelerates mutation analysis by 35.31\%, with an average 1.96\% error in mutation score.
Our results demonstrate that a trade-off can be made between mutation testing speed and mutation score error.



\end{abstract}

\begin{IEEEkeywords}
Deep Neural Network, Neuron, Mutation Analysis, Mutant, Clustering
\end{IEEEkeywords}

\input{sections/introduction}
\begin{figure*}[ht!]
    \centering
    \vspace{2mm}
    \includegraphics[scale=0.7]{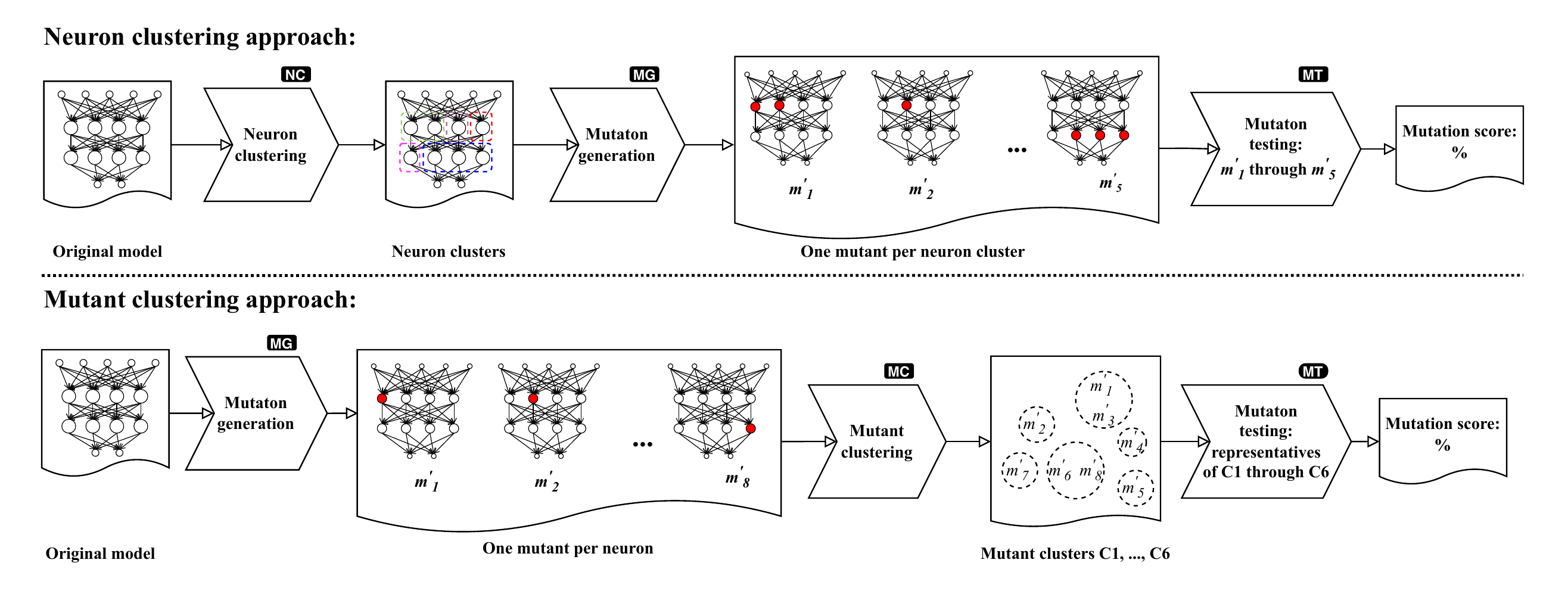}
    \caption{An overview of approaches implemented in \dmaacc: neuron clustering approach that creates fewer mutants by clustering the neurons and mutating neuron clusters instead of individual neurons (top row); mutant clustering approach that clusters the mutants and tests a representative from each cluster (bottom row). Chevrons represent processes, control flow is denoted by arrows, and document symbol denotes artifacts generated by the processes. Processes are marked with abbreviations for ease of reference in the text.}
    \label{fig:dmaacc-arch}
\end{figure*}
\input{sections/background}

\input{sections/approach}

\input{sections/experiments}

\input{sections/discussion}
\input{sections/related}
\input{sections/confuture}

\section*{Acknowledgments}
We thank the Anonymous ICST 2025 Reviewers for their insightful comments.
This research was supported by startup funding from the Department of Computer Science and Software Engineering at Auburn University.

\section*{Data Availability}
\dmaacc source code and a replication package are publicly available at~\cite{bib:replica}.

\bibliographystyle{IEEETran}
\bibliography{main}

\end{document}

%% file: misc/macros.tex
\crefformat{section}{\S#2#1#3}
\crefformat{subsection}{\S#2#1#3}
\crefformat{subsubsection}{\S#2#1#3}

\newcommand{\dmaacc}{\textsc{DeepMAACC}\xspace}

\newcommand{\ie}{\textit{i.e.}\xspace}
\newcommand{\eg}{\textit{e.g.}\xspace}
\newcommand{\etc}{\textit{etc}\xspace}
\newcommand{\perse}{\textit{per se}\xspace}
\newcommand{\ala}{\textit{à la}\xspace}
\newcommand{\cf}{\textit{c.f.}\xspace}
\newcommand{\via}{\textit{via}\xspace}
\newcommand{\vs}{\textit{vs.}\xspace}
\newcommand{\etal}{\textit{et al.}\xspace}
\newcommand{\viceversa}{\textit{vice versa}\xspace}

\newcommand{\ali}[1]{\textcolor[rgb]{1.0,0.0,0.0}{#1}}
\newcommand{\lauren}[1]{\textcolor[rgb]{0.0,0.0,1.0}{#1}}

\newtcbox{\surbox}{on line, colframe=black, colback=black, boxrule=0.5pt, arc=4pt, boxsep=0pt, left=2pt, right=2pt, top=2pt, bottom=2pt, fontupper=\color{white}\scriptsize\sffamily}

%% file: sections/introduction.tex
\section{Introduction}\label{sec:introduction}
Deep learning~\cite{bib:lecun2015deep} that is based on \textit{deep neural networks} (DNNs), has found applications in numerous fields.
These include computer vision~\cite{bib:he2016deep}, speech recognition~\cite{bib:hannun2014deep}, natural language processing~\cite{bib:goldberg2022neural}, software analysis~\cite{bib:pradel2021neural}, malware detection~\cite{bib:dahl2013large}, and more.
Among the various methods for ensuring the quality of DNNs~\cite{bib:liu2021algorithms,bib:huang2020survey}, testing is a commonly used approach~\cite{bib:zhang2020machine}.
In this method, test data points are either manually curated or automatically generated to meet specific testing requirements.
However, good performance on the test dataset does not necessarily imply the robustness and generality of a DNN model, and a systematic way for assessing the quality of the test dataset itself is needed.
This need is amplified with the growing applications of DNNs in safety and mission critical systems, such as autonomous driving~\cite{bib:bojarski2016end}, aviation~\cite{bib:naessens2017predicting}, healthcare~\cite{bib:miotto2018deep}, financial fraud detection~\cite{bib:roy2018deep}, and energy~\cite{bib:zhang2019review}.
The need for accuracy and robustness of such systems, justifies research on quality assurance of DNNs.

\textit{Mutation analysis}~\cite{bib:demillo1978hints,hamlet1977testing}, as a promising method for assessing test data quality, has recently been introduced in the context of DNNs~\cite{bib:shen2018munn,ma2018deepmutation,bib:humbatova2021deepcrime,bib:lu2022towards}.
DNN mutation analysis can either be applied to the source code constructing and training the model, aka \textit{source-level mutation analysis}, or directly to the trained model itself, aka \textit{graph-/model-level mutation analysis}.
Since their introduction, both source-level and model-level mutation analyses, have had plenty of applications~\cite{bib:wang2019adversarial,bib:hu2023muten,bib:hu2019deepmutation++,bib:lin2022robustness,bib:wang2021prioritizing,bib:hu2023aries,bib:sohn2023arachne,bib:wu2022genmunn,bib:ghanbari2023mutation,bib:riccio2021deepmetis,bib:deokuliar2023improving,bib:zohdinasab2024focused,bib:pour2021search,bib:jahangirova2021quality,bib:ghanbari2024decomposition} (see~\cref{sec:background} for more details).
Despite these opportunities, both forms of DNN mutation analysis involve generating and testing a large number of mutants, making them extremely costly processes.
For traditional programs, \ie, programs that are not exclusively based on data-driven pipelines, there is a large body of work concerning the topic of mutation analysis acceleration and their performance in various applications~\cite{bib:pizzoleto2019systematic,bib:usaola2010mutation}.
There are a few methods that have been applied in the literature for accelerating DNN mutation analysis~\cite{bib:feng2022mutation,bib:wang2023fine,bib:ghanbari2023mutation,bib:ghanbari2024decomposition,bib:li2022higher,bib:shen2021boundary}.
However, to the best of our knowledge, we still do not have any insights on the performance of techniques based on clustering in terms of mutation testing cost reduction and their cost on the mutation score.

This paper studies two methods for accelerating DNN mutation analysis that fall into the traditional categories of \textit{doing few} and \textit{doing faster} mutation analysis acceleration paradigms~\cite{offutt2001mutation}.
Specifically, we design, implement, and evaluate a system, named \dmaacc, that implements two approaches for mutation testing cost reduction through clustering for the purpose of reducing the number of generated mutants and also the number of tested mutants.
The first approach (doing few) clusters neurons within a layer, where all neurons in each of those clusters are simultaneously mutated to create a single mutant.
An entire cluster is mutated at once due to the assumption that clusters contain neurons that behave similarly, reducing redundant mutants.
This approach was recently implemented and applied for mutation-based DNN modular decomposition~\cite{bib:ghanbari2024decomposition}; however, we lack insight into its applicability for accelerating mutation analysis to assess the quality of the test dataset.
The second approach (doing faster) involves generating mutants by mutating individual neurons, as one would do in an approach like DeepMutation~\cite{ma2018deepmutation} or MuNN~\cite{bib:shen2018munn}, but the generated mutants are clustered based on the similarities of the mutated location.
\dmaacc selects a representative of each cluster and reuses the mutation testing results, \ie, whether the mutant has been killed or survived, to all other mutants in the cluster.
The intuition behind this approach is that mutations that result in similar impact are likely to behave similarly.
Although this is a well-known mutation analysis acceleration technique for traditional programs~\cite{bib:hussain2008clustering,bib:ji2009clustering,bib:yu2019clustering}, to the best of our knowledge, acceleration through mutant clustering has not been applied in the context of DNN mutation analysis yet.

In our experiments, we use two types of DNN architectures, namely, LeNet-5~\cite{lecun1998gradient} and fully-connected neural network (FCNN).
LeNet-5 architecture is a representative of the convolutional neural networks (CNNs) with no residual blocks, such as AlexNet~\cite{bib:alexnet} and VGGNet~\cite{bib:VGG16}, while FCNN architecture is a representative of simple neural networks.
These model architectures are trained on 4 popular classification datasets, \ie, MNIST~\cite{bib:deng2012mnist}, Fashion MNIST~\cite{xiao2017fashion}, Extended MNIST~\cite{cohen2017emnist}, and Kuzushiji MNIST~\cite{bib:clanuwat2018kmnist}, resulting in 8 DNN models in total.
We compare the performance of the two mutation analysis acceleration techniques to that of a \textit{vanilla method}, which simply tests all the generated mutants sequentially and exhaustively, in terms of mutation testing cost reduction as well as their cost on mutation score.
The results in our dataset of models show that, compared to the vanilla method, the first approach, \ie, neuron clustering, results in an average of 69.77\% speed-up in mutation analysis at the cost of -26.84\% average error in mutation score.
Meanwhile, the second approach accelerates mutation analysis by 35.31\%, an average, with an average mutation score error of only 1.96\%.
These results provide empirical evidence that both approaches yield significant gain in mutation testing speed with the second approach incurring less error in mutation score.
And in any potential application, it will be a matter of making a trade-off between choosing one method over another.

In summary, accelerating mutation testing, a costly step in mutation analysis, could benefit many other applications that rely on DNN mutation analysis.
This paper makes the following main contribution in this direction.
\begin{itemize}
    \item \textbf{Approach:} This paper revisits two methods for accelerating mutation testing based on clustering of neurons (to generate fewer mutants) and clustering of mutants (to test fewer mutants).
    \item \textbf{Implementation:} We have implemented the two mutation testing acceleration approaches in a publicly available, and easy to use, framework~\cite{bib:replica}, named \dmaacc.
    \item \textbf{Results:} Our results shed light on the relative performance of two mutation analysis acceleration methods by comparing them to vanilla mutation testing. Our results suggests that one of the approaches yields a higher gain in mutation testing speed, while the other incurs less error in mutation score. A replication package is available~\cite{bib:replica}.
\end{itemize}

The remainder of this paper is structured as follows. We present background information on mutation analysis in~\cref{sec:background}. The design and implementation of \dmaacc are detailed in~\cref{sec:approach}. Experimental results are discussed in~\cref{sec:experiments}, followed by an analysis of threats to validity in~\cref{sec:discussion}. Related work is reviewed in~\cref{sec:related}, and the paper concludes with insights and future directions in~\cref{sec:confuture}.

%% file: sections/background.tex
\section{Background}\label{sec:background}
\subsection{Mutation Analysis of Traditional Programs}
Mutation analysis~\cite{bib:demillo1978hints,hamlet1977testing} is a program analysis method for assessing the quality of a test suite.
This method involves generating a set of program variants, called \textit{mutants}, by mutating program elements, \eg, negating a numeric literal, using \textit{mutation operators}, aka \textit{mutators}, and running the test suite on the mutants to check if any differences between the outputs of the mutants and that of the original program is observed.
If the check is positive, the mutant is marked as \textit{killed}, otherwise it will be marked as \textit{survived}.
A mutant might survive because it is semantically equivalent to the original program, hence the name \textit{equivalent mutant}.
Test suite quality is measured by \textit{mutation score}, which is traditionally defined to be the ratio of killed mutants over non-equivalent survived mutants.
Mutation score of a test suite is assumed to be proportional to the capability of the test suite in detecting real bugs~\cite{papadakis2019mutation}. 

\subsection{Mutation Analysis of DNNs}
Although there has been attempts to reuse the existing mutation analysis tool PIT~\cite{coles2016pit} on Java-based DNN systems~\cite{chetouane2019investigating,klampfl2020mutation}, the efforts of the research community has shifted entirely toward DNN mutation analysis systems with specialized mutators.
Shen \etal~\cite{bib:shen2018munn} and Ma \etal~\cite{ma2018deepmutation} developed the first dedicated techniques for mutation analysis of DNNs.
DNN mutation can be conducted at the source level, \ie, the training program and/or the data, hence the name \textit{source-level mutation analysis}, or at the model level, \ie, the graph corresponding to the trained model, hence the name \textit{graph-level} or \textit{model-level mutation analysis}.
Both forms of mutation analysis are costly~\cite{bib:hu2019deepmutation++,bib:jahangirova2020empirical}, so there is a recent research trend in accelerating this process~\cite{bib:feng2022mutation,bib:wang2023fine,bib:ghanbari2023mutation,bib:li2022higher,bib:shen2021boundary,bib:ghanbari2024decomposition}.
The two forms differ from each other primarily in the way the mutants are generated, the former involves mutating the source and/or the training data and training the resulting mutants, while the latter directly mutates an already trained model.
The process of testing of the mutants in both forms of DNN mutation analysis are identical to each other, and one of the cost reduction methods presented in this paper is readily applicable in both contexts.
Both approaches to mutation analysis of DNNs, especially model-level mutation, has been applied in many areas, including adversarial sample detection~\cite{bib:wang2019adversarial} and generation~\cite{bib:hu2023muten}, robustness analysis~\cite{bib:hu2019deepmutation++,bib:lin2022robustness}, aiding manual labeling of test data \via prioritization of test data~\cite{bib:wang2021prioritizing}, accuracy estimation to alleviate the need for manual labeling of test data~\cite{bib:hu2023aries}, fault localization~\cite{bib:ghanbari2023mutation}, automated repair of DNNs~\cite{bib:sohn2023arachne,bib:wu2022genmunn}, modular decomposition of DNN models~\cite{bib:ghanbari2024decomposition}, and improving the quality of test dataset by generating new data points guided by mutation testing~\cite{bib:riccio2021deepmetis,bib:deokuliar2023improving,bib:zohdinasab2024focused,bib:pour2021search}.
Additionally, researchers have developed mutation analysis frameworks specialized for certain types of DNN-based systems, \eg, autonomous driving~\cite{bib:jahangirova2021quality}, as well as certain DNN flavors, such as reinforcement learning~\cite{bib:lu2022towards,thomas2025muprl}.
The cost reduction techniques implemented in \dmaacc could benefit all these application areas.

%% file: sections/approach.tex
\section{\dmaacc Design}\label{sec:approach}
\dmaacc implements two approaches to mutation analysis: (1) neuron clustering approach; (2) mutant clustering approach.
A third approach of vanilla mutation analysis, which involves sequential testing of mutants, is used as a baseline (see~\cref{sec:baseline} for more details about the vanilla approach). 
Fig.~\ref{fig:dmaacc-arch} outlines the architecture of \dmaacc and the steps involved in each of the implemented approaches.
Common to both approaches, including the vanilla baseline, is the set of mutators used to generate the mutants.
So, in what follows, we explain the set of \dmaacc mutators before explaining the mutation analysis acceleration approaches.

\subsection{\dmaacc Mutators}\label{sec:app:mutators}
\dmaacc reuses the model-level mutators from DeepMutation~\cite{ma2018deepmutation} with small modifications to eliminate the sources of randomness in them.
Removing randomness is essential for enabling us to compare the three approaches in a fair, reproducible manner in~\cref{sec:experiments}.
We would like to emphasize that \dmaacc does not make any assumption as to how the mutants are generated, and any other set of mutators can be used in practice.
So, this modification, which is done solely for the purpose of fairness and reproducibility of our experiments, does not have any impact on the applicability of our techniques.

\dmaacc creates mutants by applying a set of 3 model-level mutators that target on a single neuron or neuron cluster.
This creates 3 mutants per mutatable neuron/neuron cluster in the model, \ie, a neuron or a neuron cluster that belong to a dense or a convolutional layer.
The specific model-level mutators are as follows.
\begin{itemize}
    \item \textbf{Change Weights:} This mutator changes the weights and biases of a neuron/neuron cluster by a user-defined fraction $p$, which defaults to 0.1.
    \item \textbf{Neuron Effect Block:} This mutator blocks the effect of a neuron/neuron cluster by setting its weights and biases to zero. Therefore, this neuron/neuron cluster will not influence the DNN's classification output. It is similar to deleting the neuron/neuron cluster, as its weights no longer propagate through the rest of the model layers.
    \item \textbf{Neuron Activation Inverse:} This mutator inverses the weights and biases of a neuron/neuron cluster by multiplying the entire weight and bias vector by $-1$. It changes the sign out the output value of a neuron/neuron cluster, creating non-linear behaviors that will effect the DNN's classification output.
\end{itemize}
We have not included DeepMutation's layer-level mutators, and opted to not to study them, as those operators are known to be less effective in practice, and later works~\cite{bib:hu2019deepmutation++, bib:wang2019adversarial} did not find them useful.
However, having such mutators would not impact the operation of the cost reduction methods in \dmaacc, as they are agnostic to the mutator operator.


\subsection{Neuron Clustering Approach}\label{sec:app:nc}
Neuron clustering approach, outlined in the top row of Fig.~\ref{fig:dmaacc-arch}, begins by clustering neurons within each mutable layer based on similarities of the weights and biases between each neuron (the step marked with \surbox{NC} in Fig.~\ref{fig:dmaacc-arch}).
Agglomerative hierarchical clustering~\cite{kaufman1990finding} has been used to cluster the neurons.
In our implementation, this algorithm uses Euclidean distance~\cite{bib:wikiEuclidean} of the vectors of the weights and biases of the neurons to calculate the similarities of the neurons.
Each cluster is created with a user-specified parameter $s$ that is the average number of neurons to be put into each cluster.
For each of the $L$ mutable layers of the model $M$, we produce $\ceil*{\frac{n}{s}}$ clusters, where $n$ is the number of neurons within layer $L_i$.
In practice, the number of neurons will vary per cluster, ranging from $[1, n-\ceil*{\frac{n}{s}}+1]$, but there will always be $\ceil*{\frac{n}{s}}$ clusters for each layer.
More precisely, \dmaacc partitions the neurons per layer into clusters that have similar weights and biases before mutation generation.
The neurons per cluster parameter \textit{s} creates clusters with at most \textit{s} neurons inside.
Capping the amount of neurons within the clusters allow the parameter to equally partition clusters over every layer.
The amount of clusters will be proportional to the amount of neurons within a single mutable layer.
This parameter can generalize throughout the different mutable layers better than the alternative parameter number of clusters.
This would create a fixed set of clusters for every layer, dismissing the varying amount of neurons per layer in most models.

To generate mutants, the mutation operators are applied to each of the neuron clusters, thereby mutating all the neurons in each neuron cluster.
The motivation behind this approach relies on the assumption that the neurons that have similar weights and biases perform similar tasks in the DNN models~\cite{bib:Qi2022Patching,bib:Qi2024Reusing,bib:ghanbari2024decomposition}.
This clustering allows for less mutants to be created and tested, saving on time and storage, because instead of having 3 mutants per neuron in each mutable layer, we will have 3 mutants per neuron cluster in each mutable layer.
During mutation testing phase, the given test dataset is test the mutants sequentially so as to calculate the size of \textit{killed classes} for each mutant.
These numbers are ultimately used to calculate mutation score.

\subsection{Mutant Clustering Approach}\label{sec:app:mc}
The \textit{mutant clustering approach}, outlined in the bottom row of Fig.~\ref{fig:dmaacc-arch}, generates the mutants by mutating individual neurons. 
After creating a set of mutants, this approach of \dmaacc clusters the mutants based on the similarity of the mutated locations (the step is marked with \surbox{MC} in Fig.~\ref{fig:dmaacc-arch}).
Using Euclidean distance as the measure of similarity, \dmaacc utilizes ParHAC clustering\cite{dhulipala2022hierarchical} to cluster the mutants.
ParHAC clustering requires a complete undirected weighted graph to function, so \dmaacc constructs a graph in the following manner.
Each node in this graph denotes a mutant and the weight of each edge is the reciprocal of the Euclidean distance of tuples of the form $(l,n,w,b)$, where $l$ is the layer number, $n$ is the index of the mutated neuron in layer $l$, $w$ is the mutated weighs inlined in the tuple, and $b$ is the mutated bias.
After passing this graph to ParHAC clusterer, together with a user configurable threshold value, $p$, called linkage threshold~\cite{dhulipala2022hierarchical}, it returns the clusters.
Linkage threshold represents the minimum similarity required for two clusters to be merged, ranging from 0 (no similarity) to 1 (identical).

During mutation testing, instead of testing all of the mutants sequentially, \dmaacc retrieves a random mutant from each cluster that will act as a representative of its respective cluster.
It then uses the given dataset to compute the size of the \textit{killed classes} for the representative, which will be reused for all the mutants in the cluster represented by the selected mutant.
These numbers are then used to calculate mutation score.

The intuition behind this approach is that mutating locations close to each other in a similar manner is likely to result in the same mutation outcome, so one can reduce the number of mutants by clustering them based on the proximity of the mutation location and the similarities of the mutated weights and biases.
This is a well-known method to curtail the costs of mutation analysis in traditional programs~\cite{bib:hussain2008clustering,bib:ji2009clustering,bib:yu2019clustering}.

%% file: sections/experiments.tex
\section{Experiments}\label{sec:experiments}
\begin{figure}[t!]
    \centering
    \vspace{2mm}
        \includegraphics[scale=0.34]{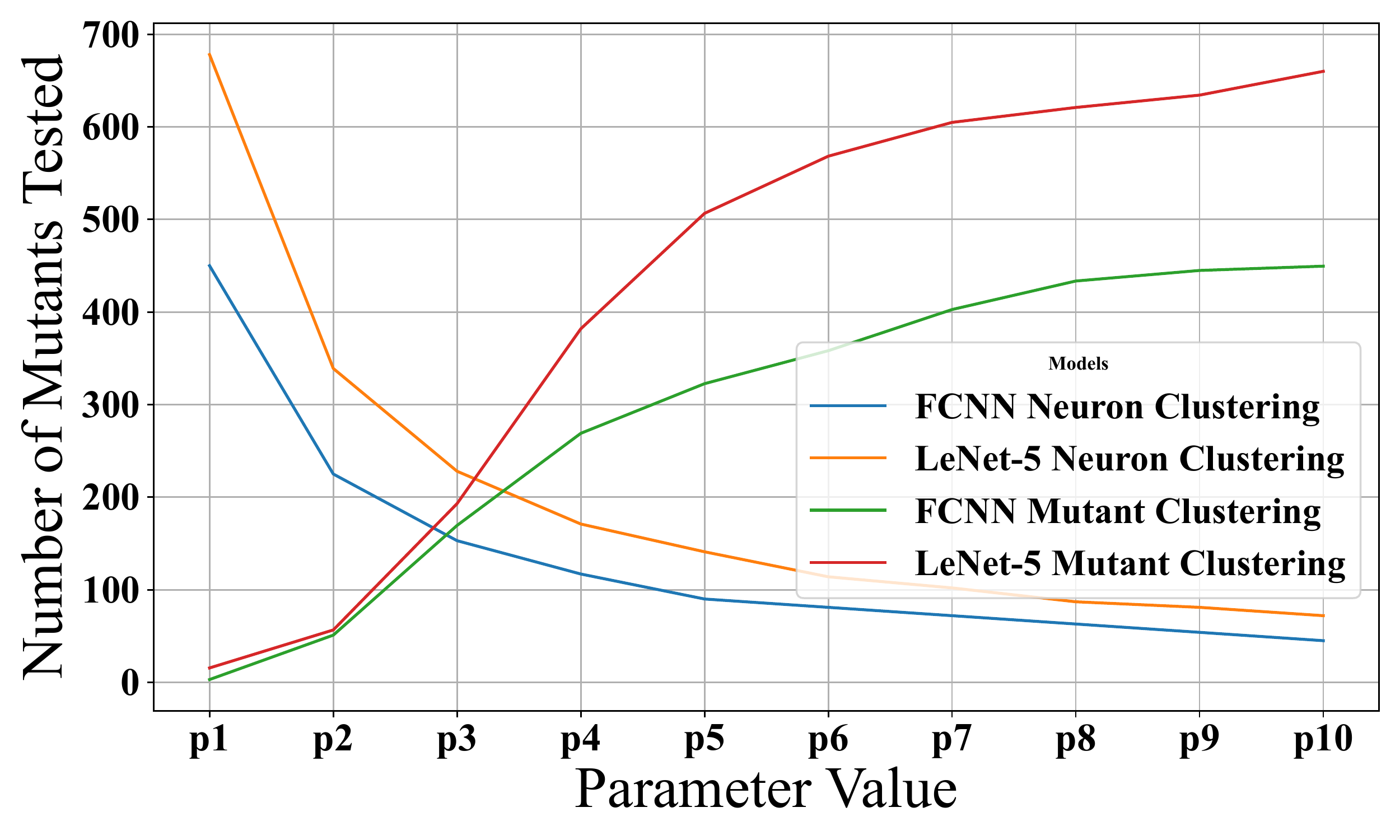}
    \caption{\textit{Number of Clusters} \vs \textit{Parameter Value} for the two different approaches used by \dmaacc. Each of the four lines represent the average \textit{Number of Mutants Tested} for a different DNN architecture and clustering approach. The parameter for Neuron Clustering is \textit{Neurons per Cluster}, having the values of \{1, 2, 3, 4, 5, 6, 7, 8, 9, 10\}. The parameter used for Mutant Clustering is \textit{ParHAC Threshold} having the values of \{0.1, 0.2, 0.3, 0.4, 0.5, 0.6, 0.7, 0.8, 0.9, 0.99\}.} 
    \label{fig:clusterplot}
\end{figure}

In this section, we apply \dmaacc on a set of DNN models to answer the following research questions (RQs).
\begin{itemize}
     \item \textbf{RQ1:} How much speedup is gained when using neuron clustering or mutant clustering versus vanilla mutation testing?
     \item \textbf{RQ2:} How much mutation score is lost when using neuron clustering or mutant clustering versus vanilla mutation testing?
     \item \textbf{RQ3:} What is the impact of non-determinism in the training process on the mutation testing speedup and mutation score of \dmaacc when using neuron clustering or mutant clustering versus vanilla mutation testing?
\end{itemize}
To study the two clustering approaches of \dmaacc, the values of their two respective parameters, namely neurons per cluster $s$ and ParHAC linkage threshold $p$, are varied.
\dmaacc is ran on a set of DNN models for each parameter.
For neuron clustering approach, $s$ is ranged over the sequence of numbers from 1 to 10, creating mutants of the neuron clusters of size 1 to 10.
For mutant clustering approach, $p$ is ranged over 0.1 to 0.99 to account for different linkage thresholds.
The reader is referred to~\cref{sec:exp:benchmark} for more details about our methodology.

\dmaacc experiments are conducted on two identical Dell Precision workstations with AMD Ryzen Threadripper @ 2.7 GHz CPU, 1 TB of RAM, and two NVIDIA RTX A6000 GPUs.
These two machines run Ubuntu 22.04.4 LTS.

\subsection{Datasets of DNN Models}\label{sec:exp:dataset}
Two types of DNN architectures are used to evaluate \dmaacc: fully-connected neural networks (FCNN) and LeNet-5\cite{lecun1998gradient}.
There are a total of 4 FCNN models and 4 LeNet-5 models that are trained on the well-known classification datasets MNIST\cite{lecun1998gradient}, Extended MNIST\cite{cohen2017emnist} (EMNIST), Fashion MNIST\cite{xiao2017fashion} (FMNIST), and Kuzushiji MNIST\cite{bib:clanuwat2018kmnist} (KMNIST).
Datasets MNIST, FMNIST, and KMNIST represent 10-class classification problems, while EMNIST is for a 26-class classification problem.
The architecture of the FCNN model consists of an input layer, a flatten layer, 3 hidden layers each with 50 ReLU-based neurons, and an output layer with softmax activation.
The LeNet-5 architecture consists of an input layer, 2 convolution layers and 3 fully-connected layers with tanh activation, and a softmax output layer.
Table~\ref{table:Architectures} presents more details about our dataset.

\subsection{Baseline: Vanilla Approach}\label{sec:baseline}
We compare the two approaches of \dmaacc to a baseline mutation analysis approach that we call \textit{vanilla approach}.
This approach follows the classic method of DNN mutation analysis: it is the same approach of mutant generation and testing employed by systems like DeepMutation~\cite{ma2018deepmutation}, where the generated mutants are tested sequentially.
During the mutation generation phase, the vanilla approach uses the 3 mutation operators, discussed in~\cref{sec:app:mutators}, to mutate each neuron within each mutable layer of the model.
This creates 3 different mutants per neuron in each mutable layer.
During mutation testing phase, the provided test dataset is used to test the mutants sequentially so as to calculate a mutation score.

\subsection{Measures}
\subsubsection{Mutation Score}
To study the impact of the mutation testing acceleration techniques on the mutation score, we use the mutation score calculation formula for DNN classifiers introduced by Ma \etal~\cite{ma2018deepmutation}.
Assume that $m$ is a $k$-class classifier and let $C=\{c_1,\dots,c_k\}$ be all the $k$ classes of the test dataset $T$, \ie, $C=\{\mbox{argmax}(t)|t\in T\}$.
Given a test data point $t\in T$, $t$ is said to kill the class $c\in C$ of a mutant $m'$, if $t$ is correctly classified as $c$ by the original model $m$, yet it is not classified as $c$ by $m'$.
Now for a set of mutants $M$, the mutation score is defined to be:
\begin{equation}\label{Mutation_Score_Formula}
    \frac{ \Sigma_{m' \in M} |\mbox{KilledClasses}(T, m')|}{|M'| \times |C|},
\end{equation}
where KilledClasses denotes the set of classes of $m'$ killed by test data in $T$.

\subsubsection{Mutation Testing Speedup}
Speedup, \ie the amount of acceleration, is calculated by dividing the difference between the average vanilla mutation testing time $t_{\text{v-avg}}$ and the average mutation testing time for neuron/mutant clustering approach, \ie, one of the two approaches implemented by \dmaacc, $t$ by the average vanilla mutation testing time $t_{\text{v-avg}}$.
It represents the percentage of reduction in mutation testing time due to the use of any of \dmaacc's mutation testing cost reduction approaches \vs the vanilla approach.
\begin{equation}\label{Speedup_Formula}
S = \frac{t_{\text{v-avg}} - t}{t_{\text{v-avg}}}
\end{equation}
All mutation testing times, \ie the time needed for testing a set of generated mutants, are measured in seconds.
\subsubsection{Mutation Score Error (Loss)}
Mutation score error, or loss, that is calculated based on the Formula~\ref{Mutation_Score_Formula}.
This is the percentage of deviation of \dmaacc mutation score from vanilla mutation score, and is calculated by dividing the difference between the average vanilla mutation testing score $s_{\text{v-avg}}$ and the average neuron/mutant clustering-based mutation testing score $s$ by the average vanilla mutation testing score $s_{\text{v-avg}}$.

\begin{equation}\label{Loss_Formula}
L = \frac{s_{\text{v-avg}} - s}{s_{\text{v-avg}}}
\end{equation}

\begin{table}[t!]
\centering
\vspace{2mm}
\caption{Dataset of DNN models used in our experiments*}
\label{table:Architectures}
\resizebox{\columnwidth}{!}{
\begin{tabular}{||c | c | r | r | r | r | r ||} 
 \hline
 \textbf{Arch.} & \textbf{Dataset} & \textbf{Model Size} & \textbf{Train \#} & \textbf{Test} \# & \textbf{Classes} & \textbf{Test Acc.}\\ [0.7ex] 
 \hline\hline
  FCNN & EMNIST & 480,383 & 124,800 & 20,800 & 26 & 0.8698 \\\hline
 FCNN & FMNIST & 477,782 & 60,000 & 10,000 & 10 & 0.8744 \\\hline
 FCNN & KMNIST & 477,782 & 60,000 & 10,000 & 10 & 0.8678 \\\hline
 FCNN & MNIST & 477,782 & 60,000 & 10,000 & 10 & 0.9661 \\\hline
 LeNet-5 & EMNIST & 190,355 & 124,800 & 20,800 & 26 & 0.9195 \\\hline
 LeNet-5 & FMNIST & 186,020 & 60,000 & 10,000 & 10 & 0.9026 \\\hline
 LeNet-5 & KMNIST & 186,020 & 60,000 & 10,000 & 10 & 0.9495 \\\hline
 LeNet-5 & MNIST & 186,020 & 60,000 & 10,000 & 10 & 0.9871 \\ [0.4ex] 
 \hline
\end{tabular}
}

\vspace{2mm}
\begin{scriptsize}
        \begin{singlespace}
        * Train \# and Test \# represent the number of data points in the train and test datasets, respectively, and Classes denotes the number of classes/labels in the test dataset.
        \end{singlespace}
        \end{scriptsize}
\end{table}

\begin{figure*}[ht!]
    \centering
    \vspace{2mm}
    \begin{tabular}{cc}
        \includegraphics[scale=0.29]{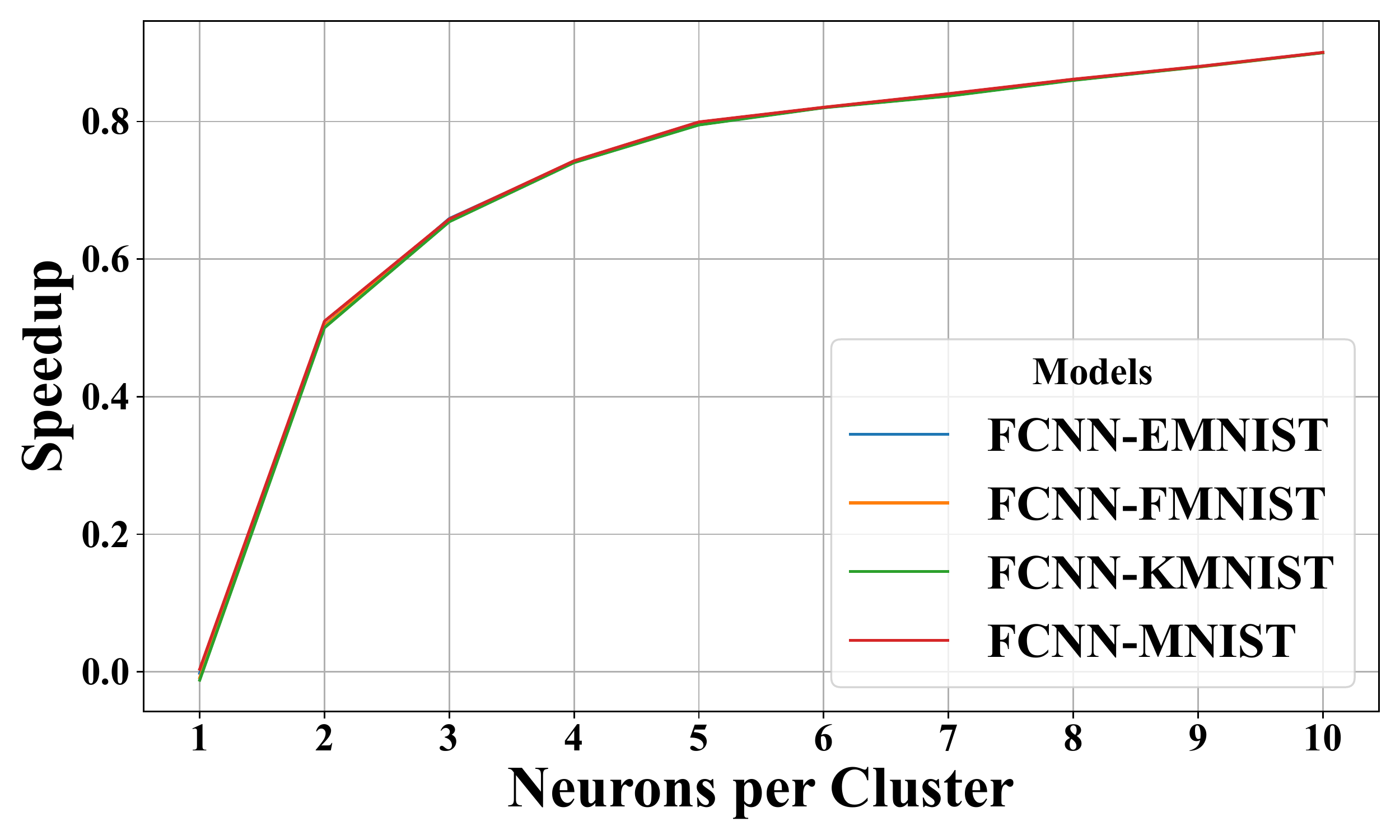} & \includegraphics[scale=0.29]{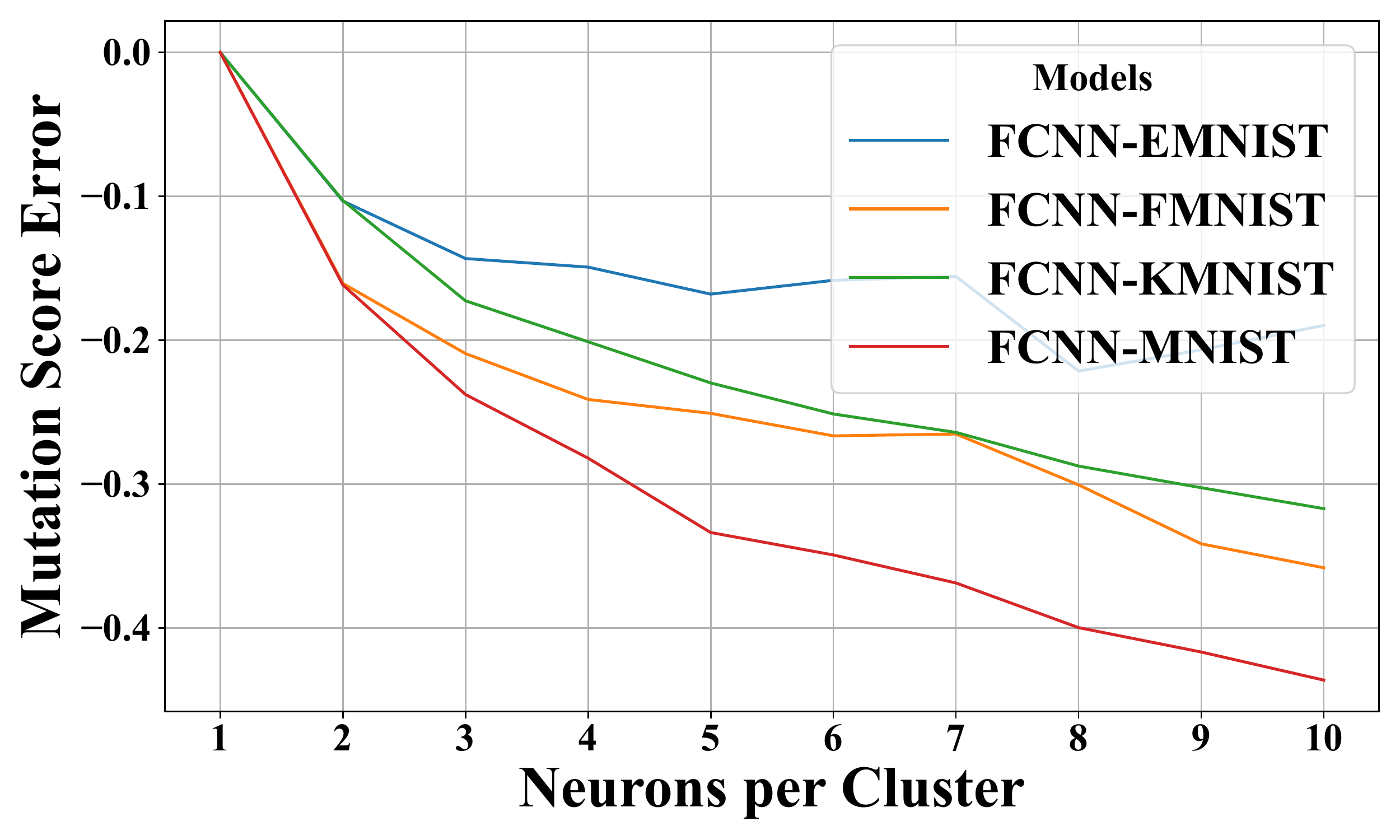} \\
        \includegraphics[scale=0.29]{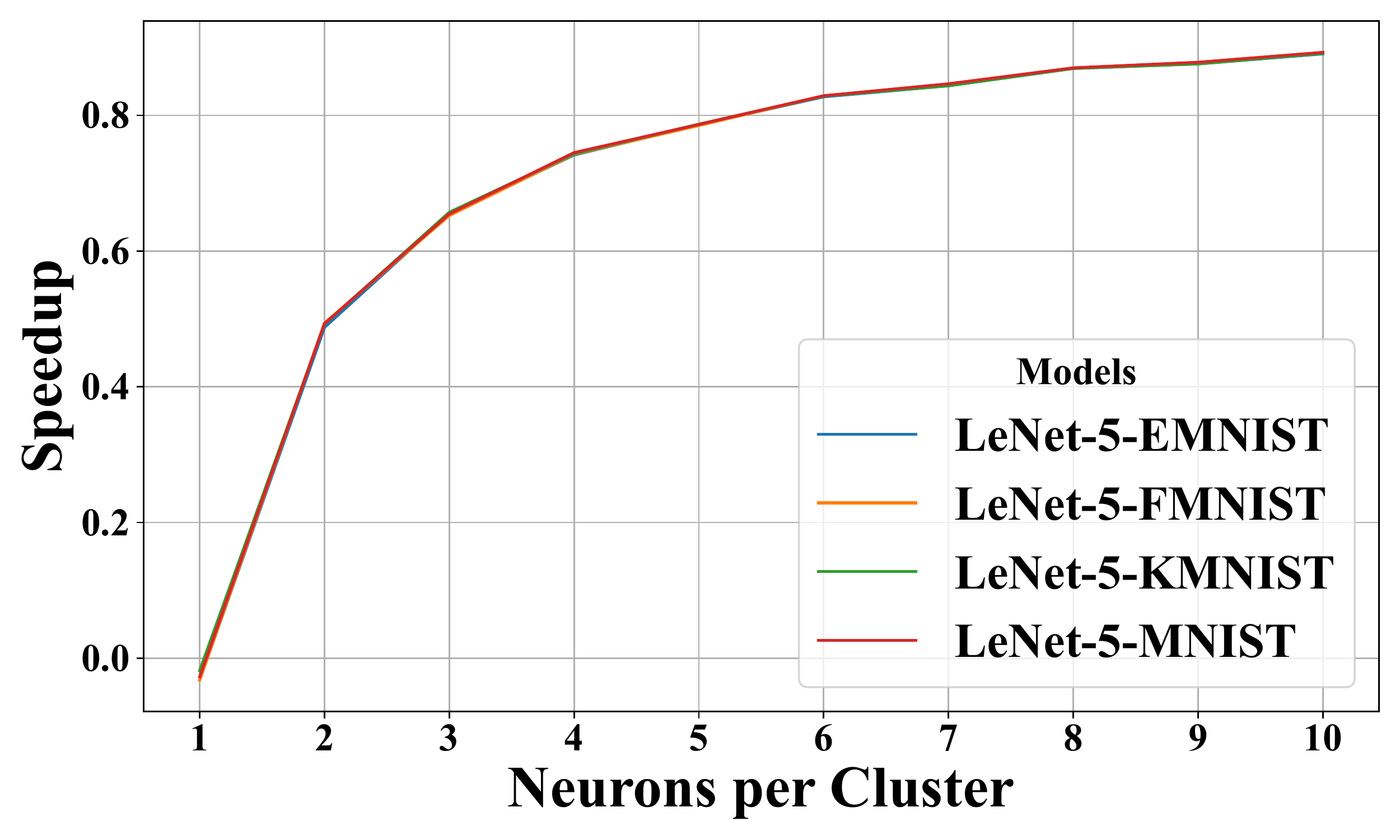} & \includegraphics[scale=0.29]{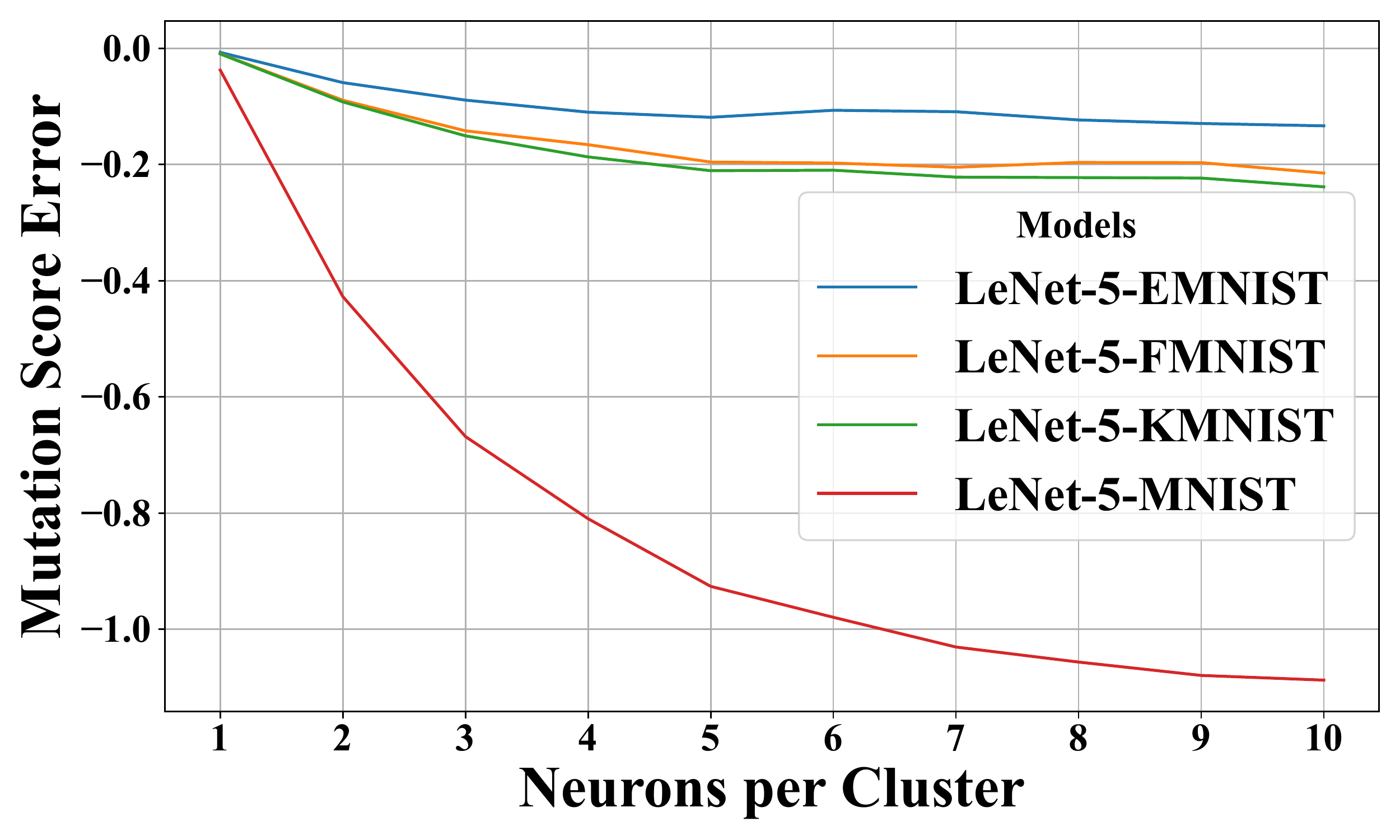} \\
        \includegraphics[scale=0.29]{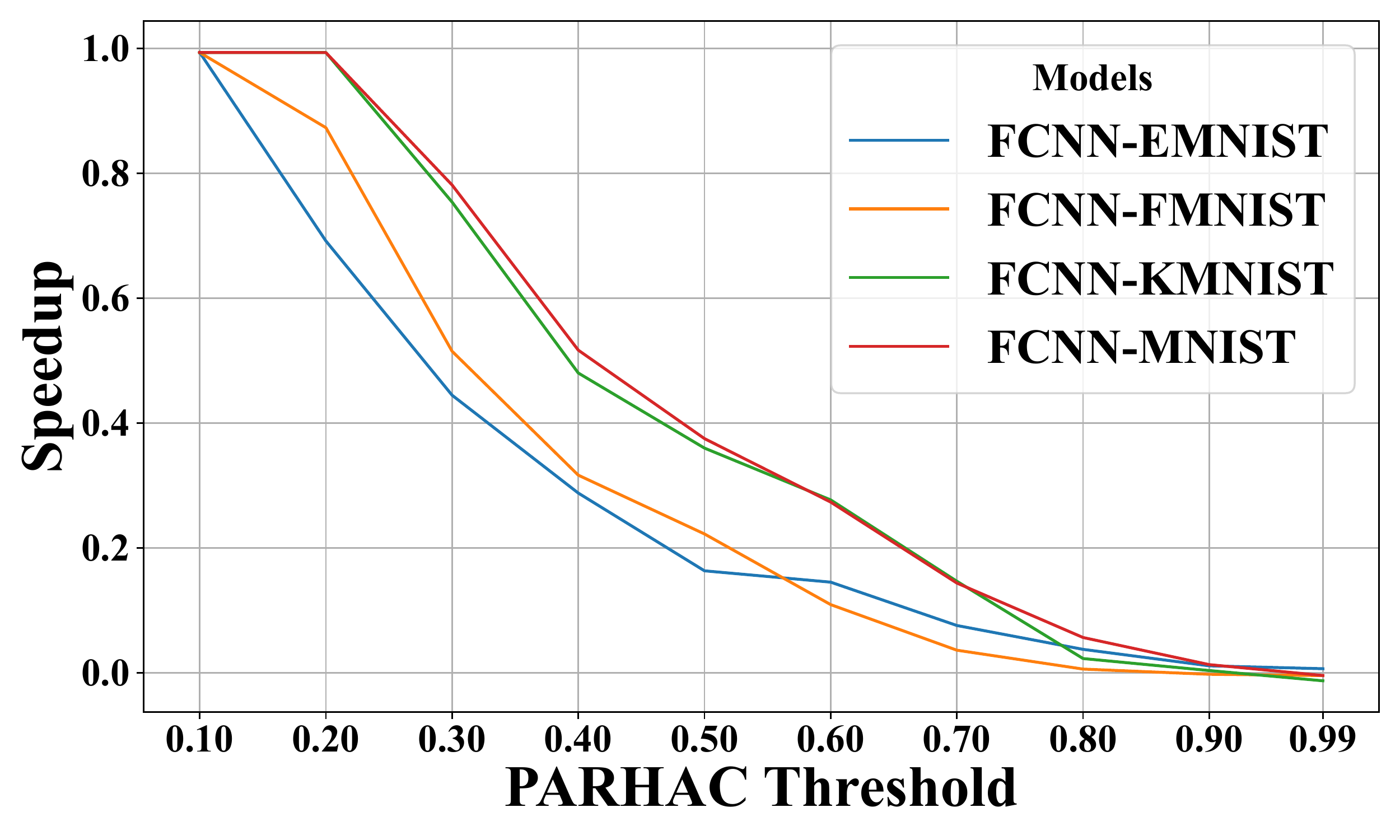} & \includegraphics[scale=0.29]{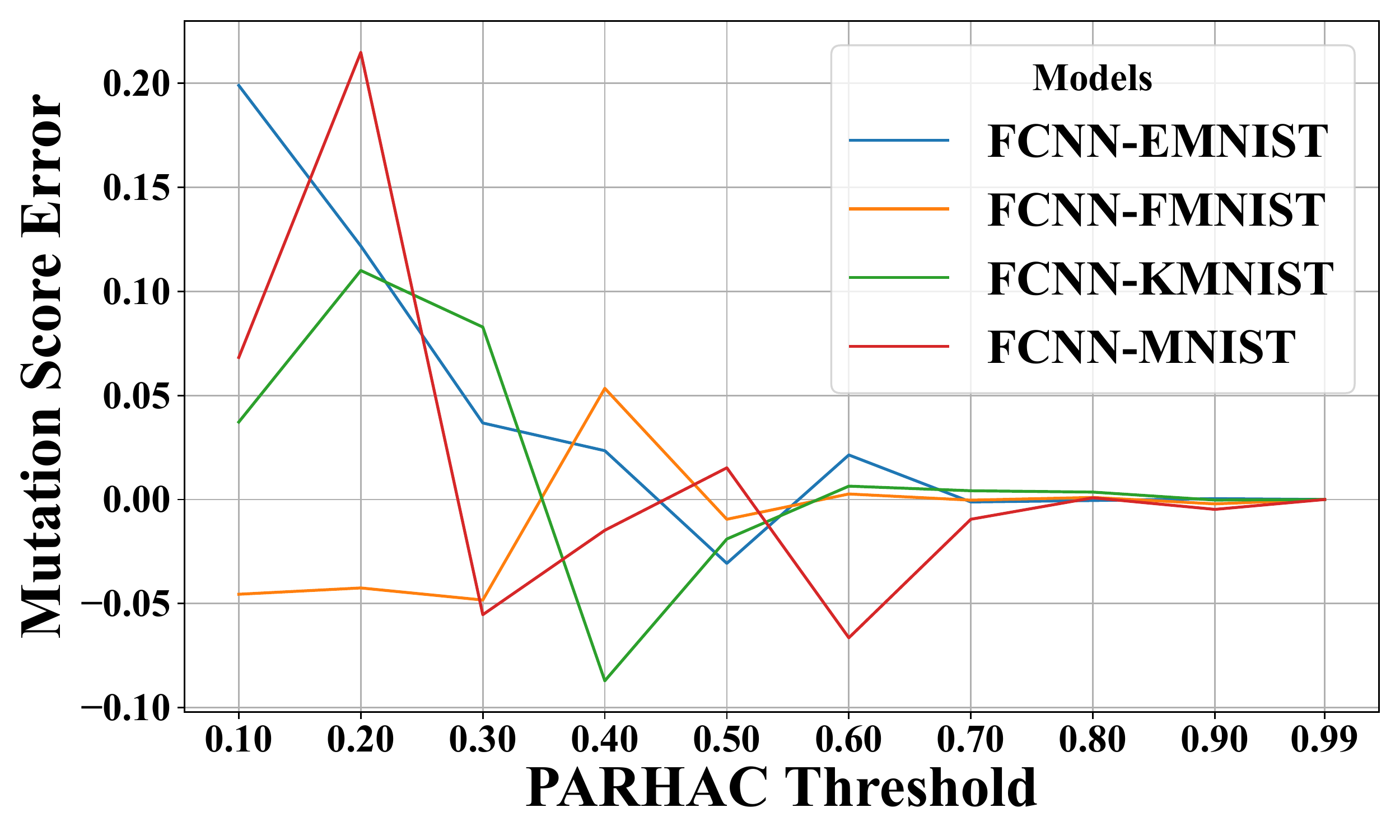} \\
        \includegraphics[scale=0.29]{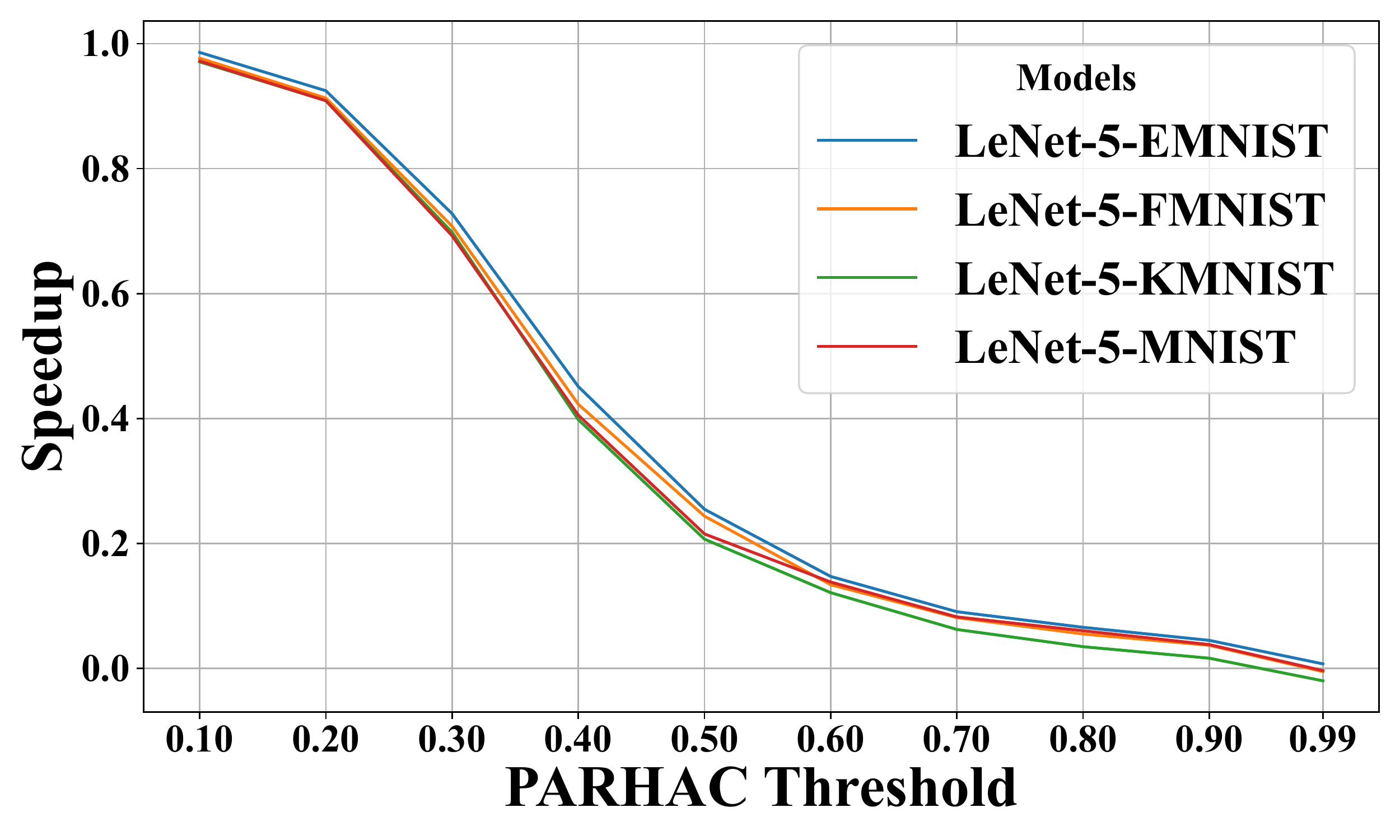} & \includegraphics[scale=0.29]{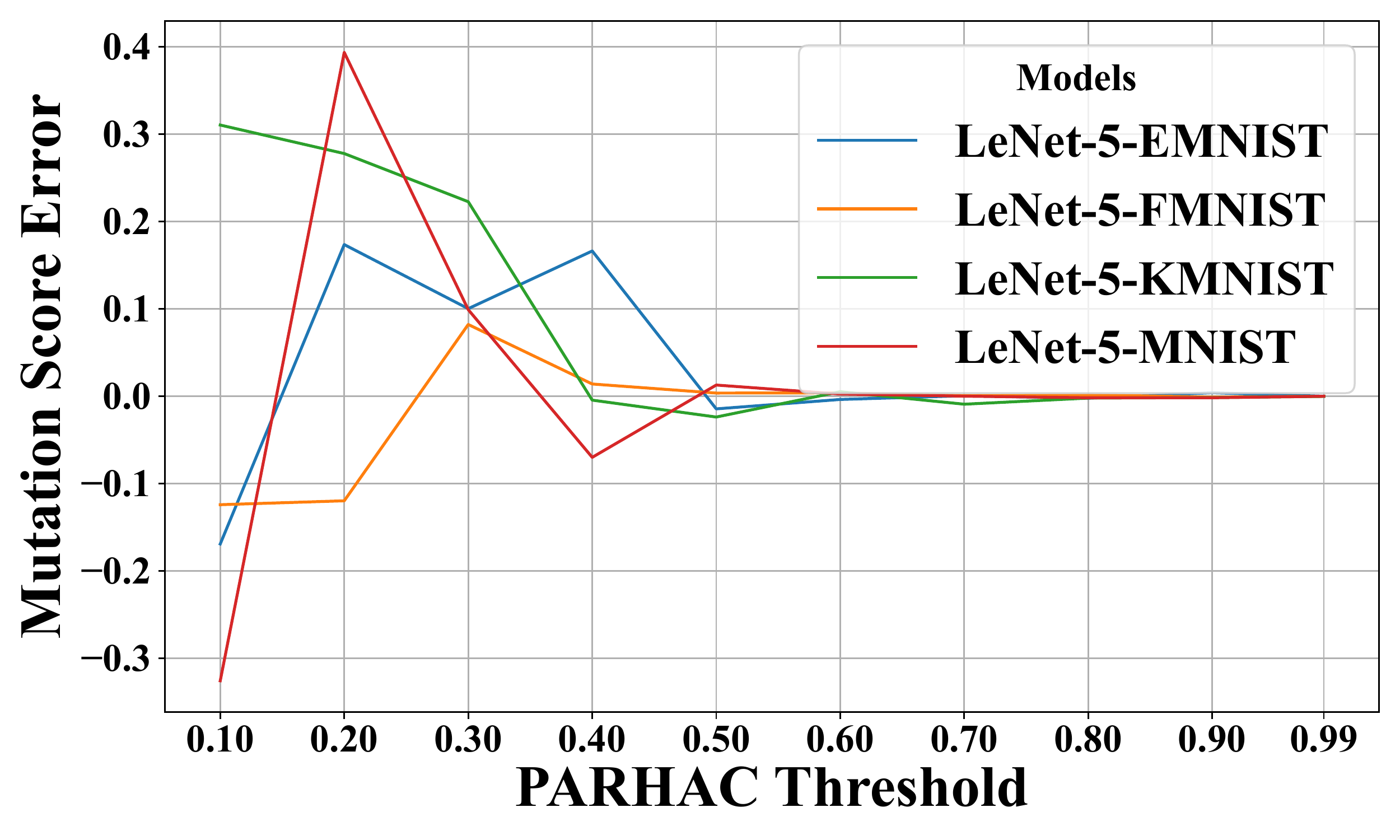}
    \end{tabular}
    \caption{From the left to right: Column 1 plots the Speedup, while Column 2 shows the Mutation Score Error metric. From the top to bottom: Rows 1 and 2 reports the results of the Neuron Clustering Approach for FCNN and LeNet-5 models, respectively, and Rows 3 and 4 present the results of the Mutant Clustering Approach for FCNN and LeNet-5 models, respectively.}
    \label{fig:speedup&mseplots}
\end{figure*}

\subsection{Method}\label{sec:exp:benchmark}
All the three approaches studied in this paper, namely neuron clustering, mutant clustering, and vanilla, have been executed multiple time per model in order to account for randomness and to align with the many executions of the two approaches due to their various parameter values.
Our neuron clustering approach is executed 6 times per a parameter value $s$ (explained in ~\cref{sec:app:nc}), which ranges over the set \{1, 2, 3, 4, 5, 6, 7, 8, 9, 10\}, while the mutant clustering approach executes 6 times per a ParHAC threshold value $p$ (explained in ~\cref{sec:app:mc}), which ranges over the set \{0.1, 0.2, 0.3, 0.4, 0.5, 0.6, 0.7, 0.8, 0.9, 0.99\}.
ParHAC threshold can only range from [0, 1.0], but we go from 0.1 to 0.99 because we do not need to test where the clusters need no similarity or where they have full similarity, \ie, they are identical.
These ten parameters for mutant clustering approach align with the ten parameters for neuron clustering approach, allowing a meaningful side-by-side comparison.
Overall, there are 1,200 executions of \dmaacc that we collected data from, resulting from 150 executions for each model: 30 runs from vanilla approach, 6 executions for each of the 10 parameters for neuron clustering approach, and 6 executions for each of the 10 parameters for mutant clustering approach. 



\begin{figure*}[ht!]
    \centering
    \vspace{2mm}
    \begin{tabular}{cc}
        \includegraphics[scale=0.29]{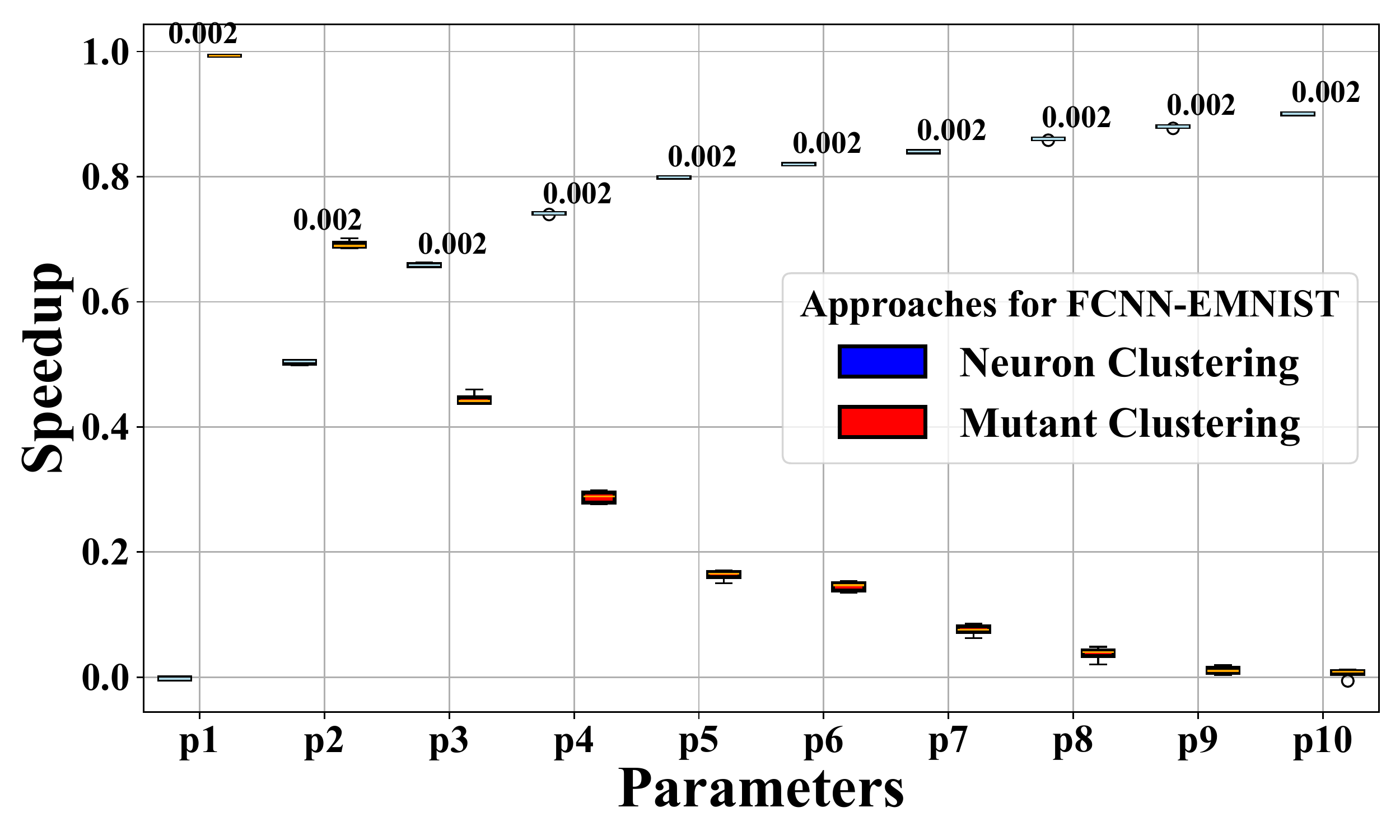} & \includegraphics[scale=0.29]{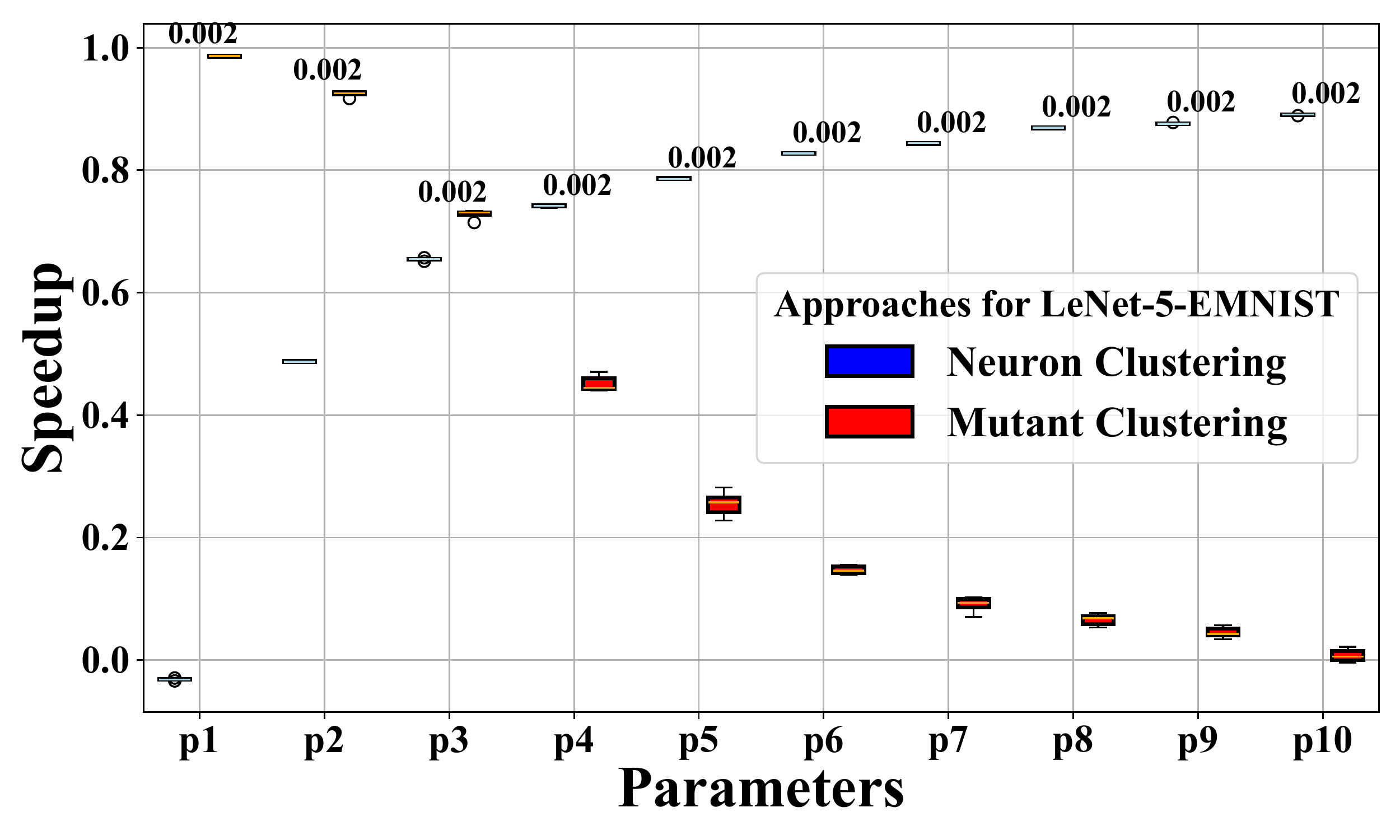} \\
        \includegraphics[scale=0.29]{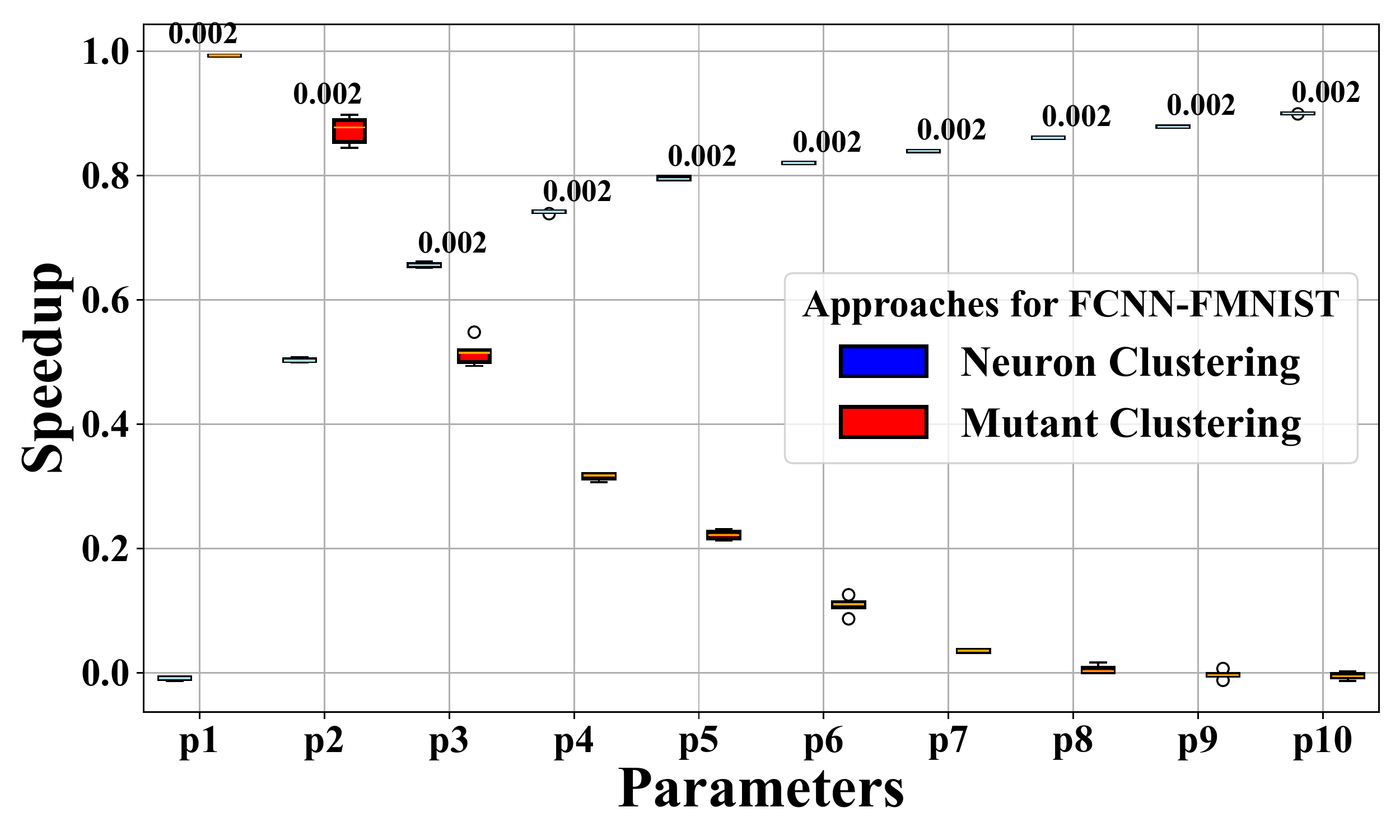} & \includegraphics[scale=0.29]{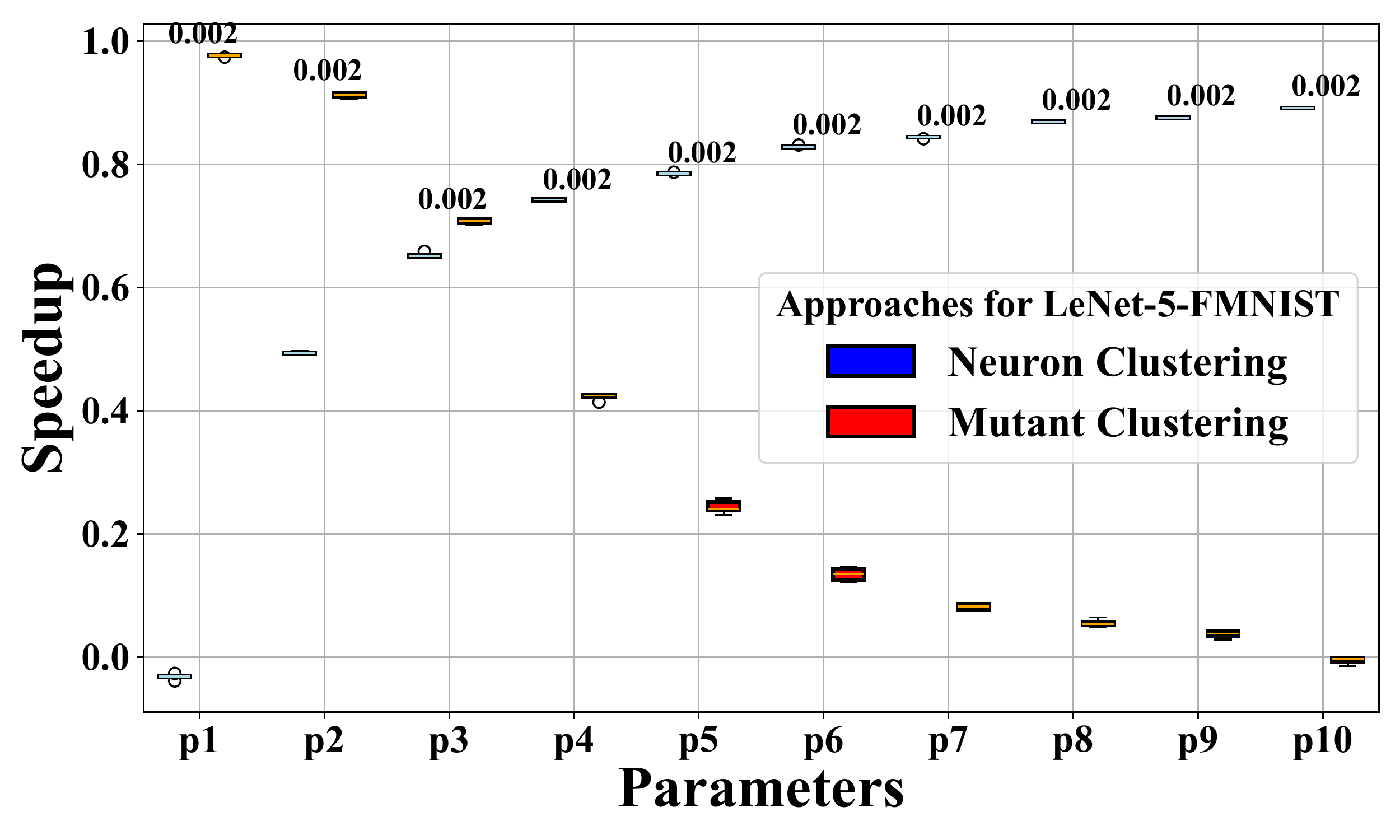} \\
        \includegraphics[scale=0.29]{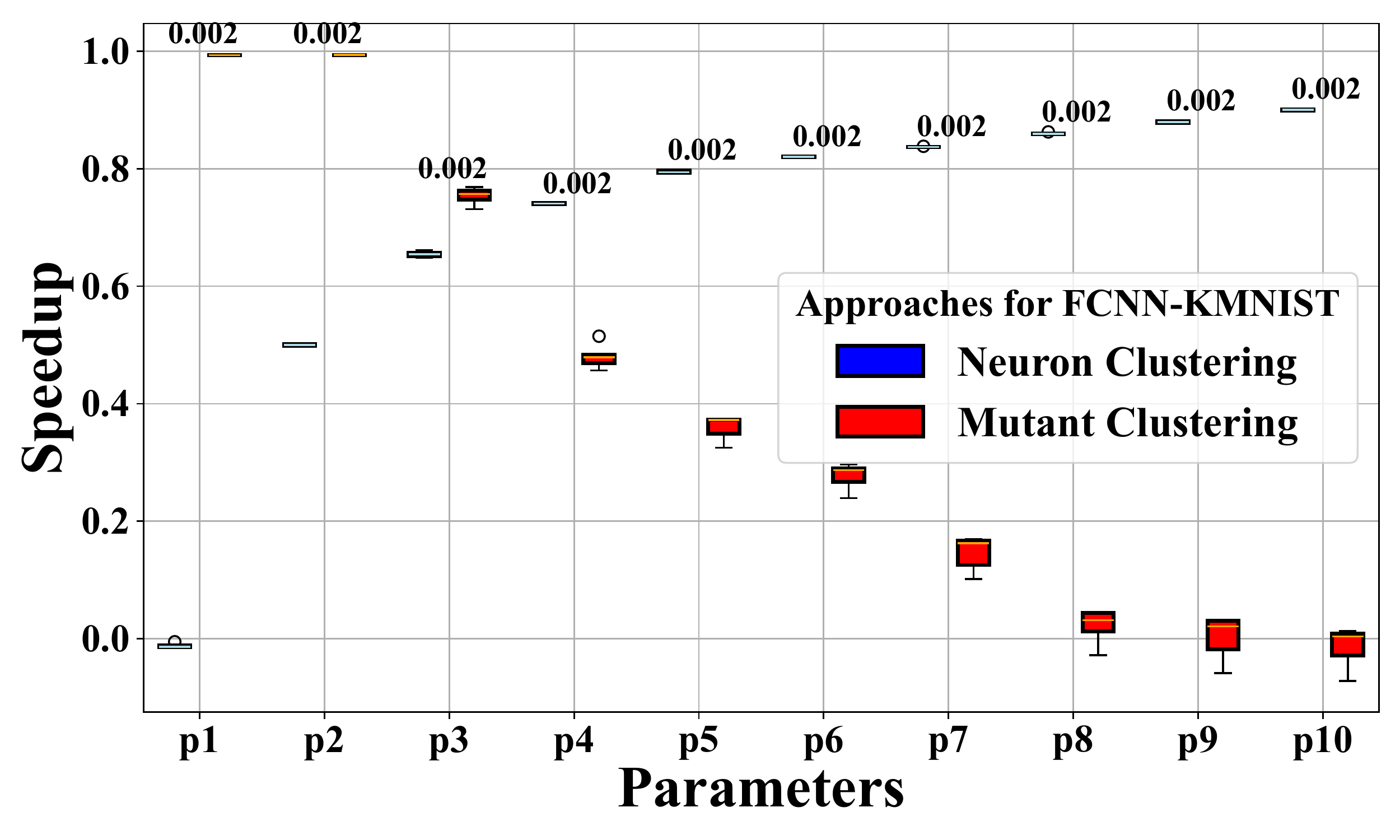} & \includegraphics[scale=0.29]{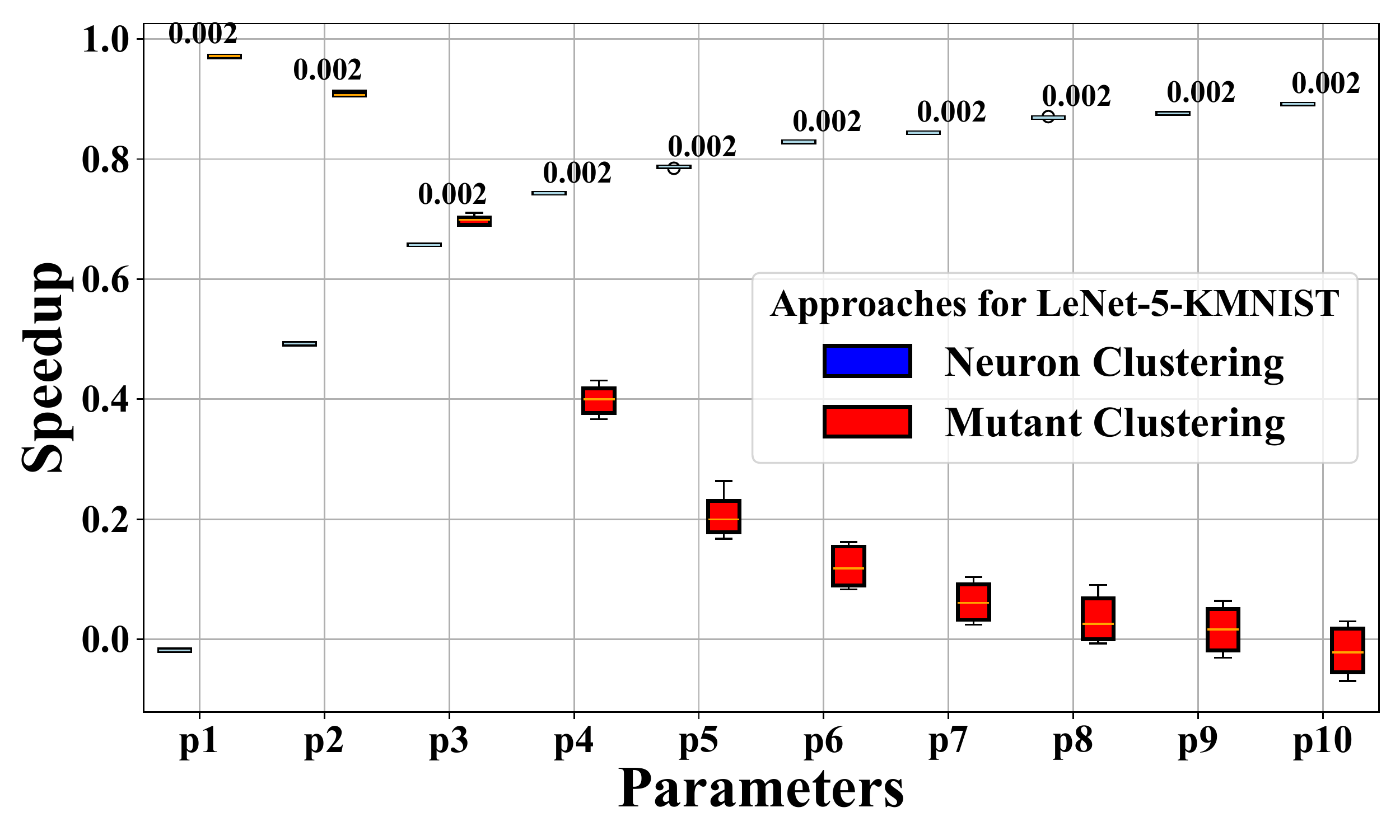} \\
        \includegraphics[scale=0.29]{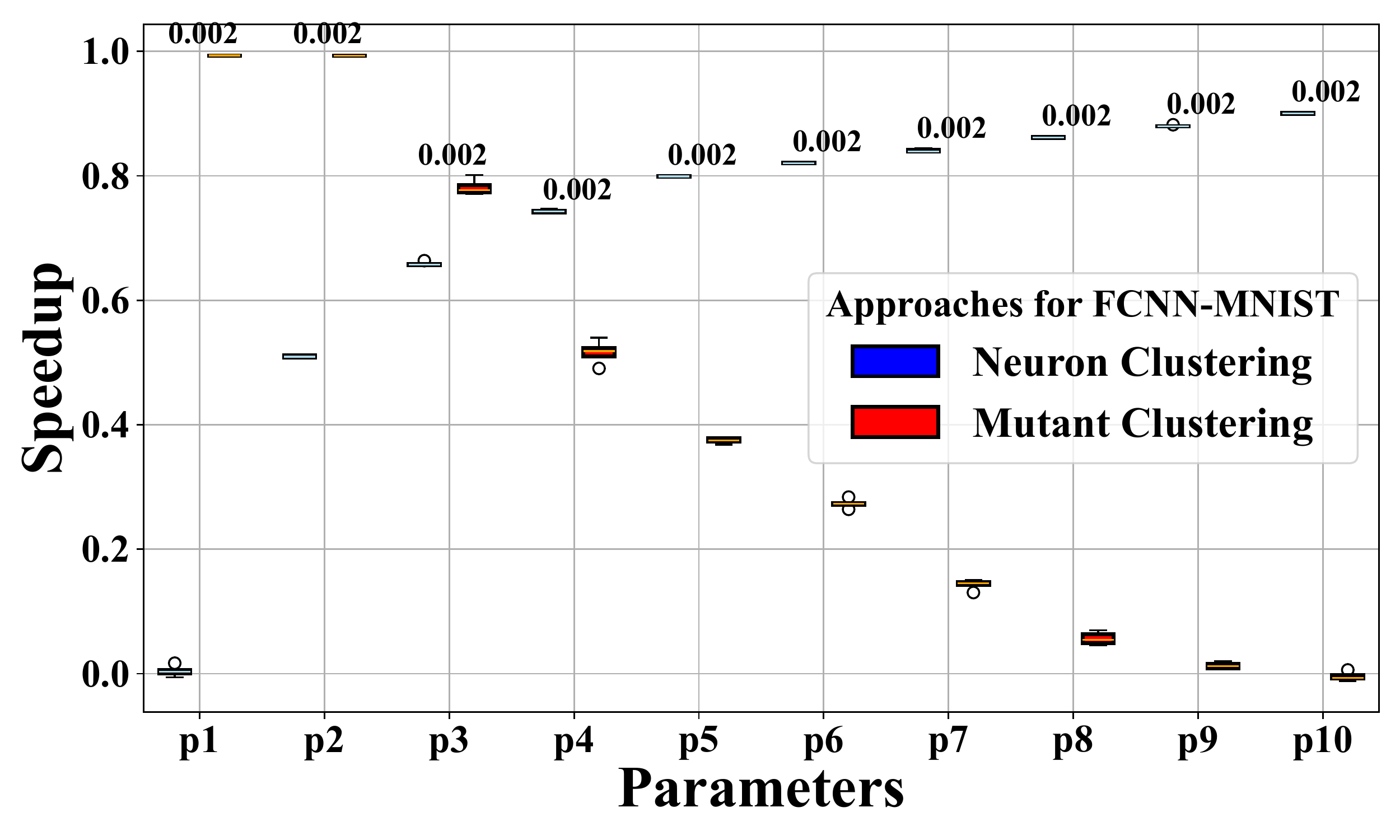} & \includegraphics[scale=0.29]{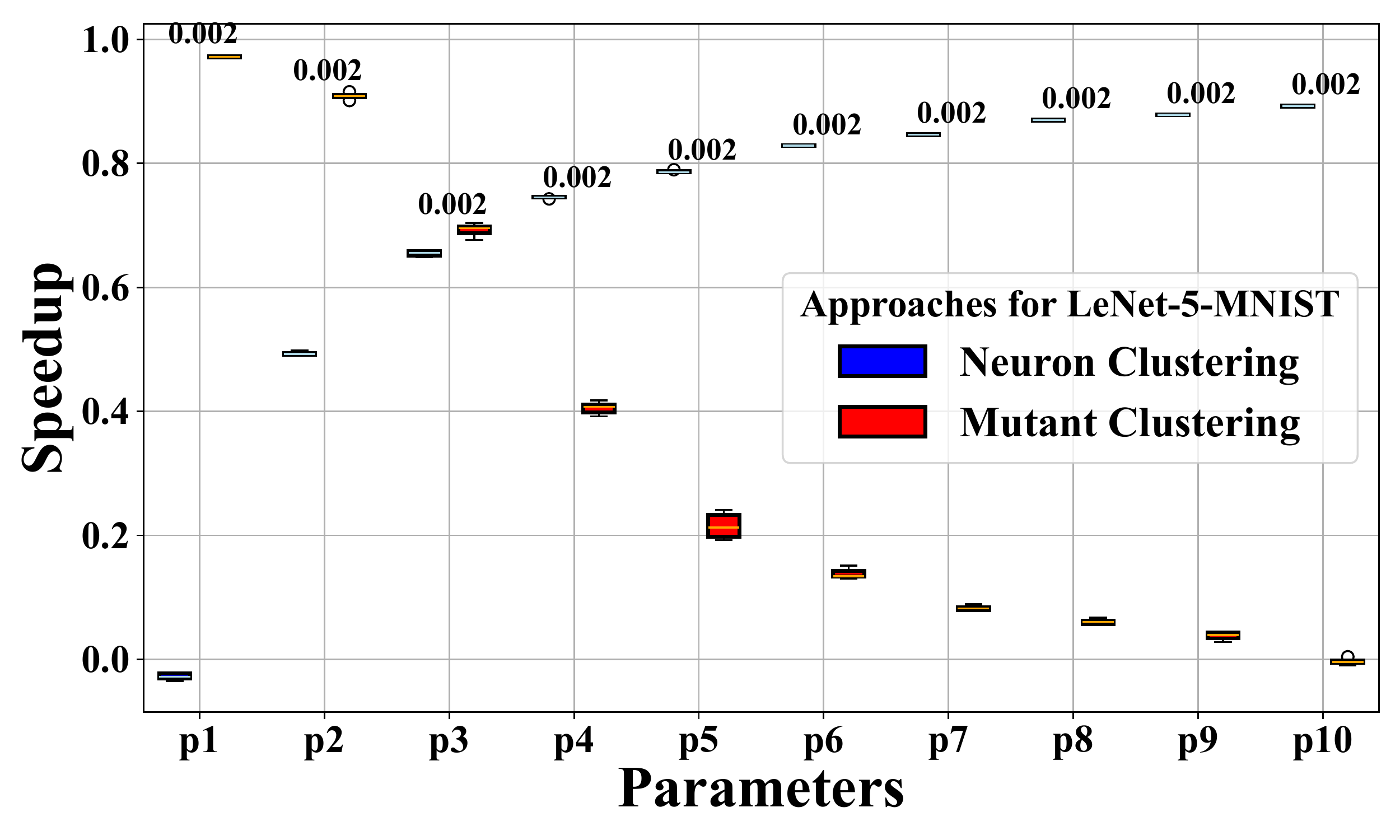}
    \end{tabular}
    \caption{Box plot pairs for mutation testing Speedup for the 8 different models. In each pair, blue box-and-whisker, on the left, represents speedup for Neuron Clustering Approach, and the red box-and-whisker, on the right, represents that of Mutant Clustering Approach. Each pair of box-and-whiskers in each plot is annotated with the $p$-value obtained \via Mann-Whitney U-Test. The parameter used for Neuron Clustering is \textit{Neurons per Cluster}, ranging over \{1, 2, 3, 4, 5, 6, 7, 8, 9, 10\}, while the parameter used for Mutant Clustering is \textit{linkage threshold}, ranging over \{0.1, 0.2, 0.3, 0.4, 0.5, 0.6, 0.7, 0.8, 0.9, 0.99\}. In column one, from top to bottom, the model's \textit{SpeedUp} that is shown are FCNN-EMNIST, FCNN-FMNIST, FCNN-KMNIST, FCNN-MNIST, respectively. In column two, from top to bottom, the model's \textit{SpeedUp} that is reported is  LeNet-5-EMNIST, LeNet-5-FMNIST, LeNet-5-KMNIST, LeNet-5-MNIST, respectively.} 
    \label{fig:box-plots}
\end{figure*}

\subsection{Results} \label{sec:exp:results}
Fig.~\ref{fig:clusterplot} plots the average number of mutants tested for each parameter for both the FCNN and LeNet-5 models using Neuron Clustering or Mutant Clustering.
This plot shows the difference in number of tested mutants, which can help explain the behavior and use of the different parameters.
For Neuron Clustering, the \textit{Number of Mutants Tested} decreases as the parameter increases.
This is expected, as the more Neurons per Cluster there are, the less clusters are required to hold all neurons.
Larger clusters results in less clusters and therefore fewer mutants as there are 3 mutants created for each cluster.
For Mutant Clustering, the \textit{Number of Mutants Tested} increases as the parameter increases.
This is expected, as the similarity threshold increases, the ParHAC clusterer requires neurons to hold more similar weights to cluster them.
A small similarity threshold will create large clusters, resulting in less mutants to test.
A large similarity threshold requires the neurons to be nearly identical, which results in smaller clusters, increasing the number of mutants to be tested.
These patterns also show up in later figures, so this explanation seems to remain consistent.

Fig.~\ref{fig:speedup&mseplots} shows the \textit{Speedup} (Column 1) and \textit{Mutation Score Error} (Column 2) of our different models and clustering approaches over multiple executions as described in \cref{sec:exp:benchmark}. We use the four FCNN models and four LeNet-5 models described earlier in this section. The plots in this figure are separated by DNN architecture type, FCNN (Row 1 and 3) and LeNet-5 (Row 2 and 4). Each of the colored lines in each plot correspond to different models. Blue lines correspond to a model (FCNN or LeNet-5) trained on the EMNIST dataset, yellow corresponds to a model trained on FMNIST, green corresponds to a model trained on KMNIST, and red corresponds to a model trained on MNIST. 

Fig.~\ref{fig:box-plots} shows the \textit{Speedup} for both clustering approaches across the 8 models represented by the 8 plots. Each parameter has a pair of box-and-whisker plots. The blue box-and-whisker plot on the left represents neuron clustering \textit{Speedup} while the red box-and-whisker plot on the right represents mutant clustering \textit{Speedup}. So, the parameters on the y-axes for neuron clustering is \textit{Neurons per Cluster} and \textit{ParHAC Threshold} for mutant clustering. The annotation above each pair is the p-value obtained via Mann-Whitney U-Test. In column one, from top to bottom, the models represented are FCNN-EMNIST, FCNN-FMNIST, LeNet-5-EMNIST, and LeNet-5-FMNIST. In column two, from top to bottom, the models represented are FCNN-KMNIST, FCNN-MNIST, LeNet-5-KMNIST, LeNet-5-MNIST. 

\subsubsection{Answering RQ1} This research question investigates the effectiveness in Speedup of \dmaacc through varying parameter values and models for each approach of the mutation testing. We will answer this for our neuron clustering approach and our mutant clustering approach separately. 

In the first column of Fig.~\ref{fig:speedup&mseplots}, we will observe the first two rows of plots to answer this question for neuron clustering. In these two plots, it is shown that the \textit{Speedup} for all the models is nearly identical at the separate parameter values.
The \textit{Speedup} does continually increase in a non-linear fashion, where an increase in \textit{Speedup} indicates a decrease in mutation testing time. A value of 1 neuron per cluster causes the \dmaacc mutation analysis to operate similarly to the vanilla approach because each neuron will be in a cluster on its own. So, the same number of mutants will be created. This leads to the initial \textit{Speedup} value of 0.0 since the vanilla and neuron clustering approaches are the two being compared. The graph suggests that the \textit{Speedup} will plateau at a value of 1.0. We can explain the similarity in the graphs for each model across both DNN architectures using Fig.~\ref{fig:clusterplot} because the reduction in tested mutants should be the same for each parameter. We can see the plotted lines for FCNN Neuron Clustering and LeNet-5 Neuron Clustering have similar slopes, but differ only in height, which is explained by the larger number of mutable neurons in a LeNet-5 model. Overall, we can observe a trend that neuron clustering increases the \textit{Speedup} for the mutation testing of a model as the parameter increases. FCNN models obtain an average 69.95\% \textit{Speedup} while LeNet-5 models obtain an average 69.59\% \textit{Speedup.}

In the first column of Fig.~\ref{fig:speedup&mseplots}, we will use the last two rows of plots to answer this question for mutant clustering. In the top plot, there is some deviation in the exact plotted lines, but the overall trend is similar. We observe an opposite trend than that of the neuron clustering approach, which can be explained by observing Fig.~\ref{fig:clusterplot}. In that figure, there is a positive slope that is similar to the inverse of the plot we are looking at in Fig.~\ref{fig:speedup&mseplots}. Testing more mutants takes a longer time, so as the number of mutants tested increases, the \textit{Speedup} decreases. However, the line in Fig.~\ref{fig:clusterplot} is the average of all FCNN models, so the deviation in \textit{Speedup} for Fig.~\ref{fig:speedup&mseplots} is likely explained by a deviation in size of mutant clusters, which would be due to the inherent behavior of ParHAC clustering. In the second plot, the plotted lines are almost identical, suggesting that there is less deviation in the number of clusters for this model type. The four different model's plotted lines have a negative trend, decreasing as the parameter increases. When the \textit{ParHAC Threshold} reaches 0.99, \dmaacc's mutant clustering approach should cause the mutation testing to operate similarly to the vanilla approach. The \textit{Speedup} approaches a value of 0.0 since 0.99 is approaching 1.0. FCNN models obtain an average 35.19\% \textit{Speedup} while LeNet-5 models obtain an average 35.43\% \textit{Speedup.}

\subsubsection{Answering RQ2} This research question investigates the effectiveness in Mutation Score Error of \dmaacc through varying parameter values and models for each approach. We will answer this for our neuron clustering approach and our mutant clustering approach separately. 

For neuron clustering, we test 10 different \textit{Neurons per Cluster} values for the number of neurons in each cluster \textit{n} to see the effects on mutation testing time. In Fig.~\ref{fig:speedup&mseplots}, from the second column we will use the first two rows of plots to answer this question. The first plot shows only FCNN models while the second plot shows LeNet-5 models. In both plots, we can observe the general negative trend of \textit{Mutation Score Error} as the \textit{Neurons per Cluster} increase. A negative \textit{Mutation Score Error} means that the mutation scores resulting from the \dmaacc neuron clustering approach are higher than that of the vanilla approach. As a model's mutation score approaches 1, more classes are killed by each mutant. Ideally, the \textit{Mutation Score Error} will near 0, as we do not want the mutation analysis results to change at all, as a larger change in mutation score indicates a larger change in model behavior.

\dmaacc's goal is to accelerate mutation analysis without altering the performance. There is an obvious outlier for LeNet-5-MNIST in the second plot. The reason for this large jump in error will need to be explored in future works, as the behavior does not reflect any other part of the analysis. However, it seems that the LeNet-5-MNIST and FCNN-MNIST both have worse trends than these models trained on the datasets FMNIST, KMNIST, and EMNIST. Having that worse trend may align with the MNIST dataset being simpler than the other tested datasets. So, this trend shows up for both the FCNN-MNIST and LeNet-5-MNIST models for neuron clustering, which could suggest that the neuron clustering method works better the more complex the model it. In all other models across both plots, the declining trend is similar, with deviations around 0.1 as the parameter values increase. This trend is expected, as the clusters grow larger, more neurons are mutated during the production of one mutant, understandably producing a larger \textit{Mutation Score Error}. As mentioned before, the \textit{Mutation Score Error} would near 0.0 because these neurons have been clustered with the idea that they control the same behaviors of the model. So, ideally, no matter the amount of neurons we mutate, we are altering the same behavior and the same classes would be killed for each cluster as an individual neuron mutant would kill. Overall, we can observe a trend that neuron clustering increases the \textit{Mutation Score Error} in the negative direction for the mutation testing of a model as the \textit{Neurons per Cluster} increase.

For mutant clustering, we test 10 different parameter values for the ParHAC linkage threshold used to cluster mutants to see the effects on mutation testing time. In the second column, we will use the last two rows of plots to answer this question. In both plots, we witness some deviation in the beginning \textit{ParHAC Thresholds}. This is due to the randomness of the mutant cluster representative choice. In Fig.~\ref{fig:clusterplot} we can see that there is a small number of mutants tested in the beginning \textit{ParHAC Thresholds}, which means there were large clusters. When a random representative is chosen from a large cluster, it is reasonable that the mutation score error has high deviations. \dmaacc strives to group behaviorally similar mutants in the same cluster, but small similarity values will cause groups of differently behaving clusters. As the \textit{ParHAC Threshold} increases, we can see that the \textit{Mutation Score Error} stabilizes. For LeNet-5 models, there is less deviation for the different models, and the \textit{Mutation Score Error} seems to stabilize earlier. The FCNN models have more deviation and stabilize at a later \textit{Mutation Score Error}. 

With a goal of mutation testing acceleration, we can observe that by applying neuron clustering to FCNN models, we obtain an average of 69.95\% \textit{Speedup} with an average -22.52\% \textit{Mutation Score Error}. LeNet-5 models have an average of 69.59\% \textit{Speedup} with an average -31.17\% \textit{Mutation Score Error}. Applying mutant clustering to mutation analysis for FCNN models can produce an average \textit{Speedup} of 35.19\% with an average 1.41\% \textit{Mutation Score Error} and an average of 35.43\% \textit{Speedup} with an average 2.50\% \textit{Mutation Score Error} when applied to LeNet-5 models. For the mutation clustering approach, the deviations cancel out, causing the average \textit{Mutation Score Error} to be minimal.

\subsubsection{Answering RQ3} This research question investigates the effects of non-determinism in the performance of \dmaacc for \textit{Speedup} and \textit{Mutation Score Error}. Non-determinism is present in the form of model training randomness, killed classes randomness, and the random choice of a cluster representative in mutation clustering in \dmaacc's executions. Multiple executions were run for each model to mitigate some of these non-deterministic effects.

We can observe that the \textit{Speedup} is not random by assesing Fig.~\ref{fig:box-plots}. These box-and-whisker plots show \dmaacc's \textit{Speedup} performance and deviations through varying parameter values and models for each approach. The box-and-whisker plot pairs show the mutation testing \textit{Speedup} for the neuron clustering and mutation clustering approaches. The \textit{Speedup} is shown to have opposite slopes for the two different clustering approaches, suggesting that they are nearly independent of a possible random chance to obtain these values. To confirm this visual assessment, a Mann-Whitney U-Test \cite{enwiki:1248746087} has been conducted over all the pairs of series of values for the different box plots, with the null hypothesis being that there is no statistically significant difference between the speedup values of neuron clustering vs mutant clustering. The alternate hypothesis is that the difference in the values are statistically significant. The p-values obtained from the test are shown as annotations above the pairs of box-and-whiskers in Fig.~\ref{fig:box-plots}. As we can see, almost all of the values are lower than 0.05, rejecting the null-hypothesis with a 95\% confidence. From this observation, we can conclude that \dmaacc tends to produce a similar or identical amount of mutants through every execution for a single parameter.

To observe the randomness in \textit{Mutation Score Error}, we can view column 2 Fig.~\ref{fig:speedup&mseplots}. For neuron clustering in the top two plots of the right column, we can observe non-determinism's presence in the slight deviations of the graph. The jagged output could also be an inherent characteristic of the models, but we can assume it is a result of randomness present in the training of the model and the model's classification of the different classes. However, in mutant clustering, seen in the bottom two plots of the right column, we can observe a more harsh result of randomness for the smaller parameter values. This large random effect results from the mutant clustering's random approach to selecting a cluster representative. When a large cluster is created from a small similarity threshold, the grouped mutants will not all behave as similar as wanted. \dmaacc strives to group behaviorally similar mutants in the same cluster, so that when a representative of the mutant is killed (or survived), all the mutants within that cluster can be safely marked as killed (or survived) without explicitly testing them. As the \textit{ParHAC Threshold} increases, the \textit{Mutation Score Error} lines stabilize, meaning that \dmaacc is no longer choosing random representatives that do not successfully represent the entire cluster. This suggests that these \textit{ParHAC Threshold} values produce quality clusters that behave similarly. Utilizing a small similarity threshold produces unwanted outcomes based on these plots and explanations. As the model is suggestedly random, the randomness cancels itself out when calculating the mean \textit{Mutation Score Error} for the DNN architectures' plots. For mutant clustering on FCNN models, the average \textit{Mutation Score Error} is 1.41\% while the average is 2.50\% for LeNet-5 models.

Overall, it is seen that neuron clustering obtains larger \textit{Speedup} values at the cost of a larger \textit{Mutation Score Error}. Mutant clustering obtains lesser \textit{Speedup} values, but nearly mitigates the \textit{Mutation Score Error}, suggesting that this method chooses quality clusters that behave similarly.

%% file: sections/discussion.tex
\section{Discussion}\label{sec:discussion}
Speeding up DNN mutation analysis could benefit plenty of application areas that depend on it.
We presented two approaches with promising results in reducing the costs of DNN mutation testing.
Both approaches generally decrease the mutation testing time, so optimal parameters and metrics that can balance the trade-offs between \textit{Speedup} and \textit{Mutation Score Error} would benefit their further research immensely.
\dmaacc will be open-sourced, so the research community can apply and improve its clustering approaches.




\subsection{Threats to Validity}



Like any other research work with empirical results, there are potential threats to validity to our work.
The generalizability of our results is limited since we do not have a complete representative sample of all DNN models and datasets to test our clustering approaches with. 
We evaluated \dmaacc on only two DNN architectures (LeNet-5 and FCNN) and four classification datasets.
While such a collection of model architectures might include modern models like AlexNet and VGGNet, they are not fully representative of the wide variety of DNN models and tasks in practice.
Further evaluation on additional architectures (e.g. ResNet, Transformers) and datasets from diverse domains would strengthen the external validity of our findings.
There is also a need for expansion of the set of tested mutant operators. However, there are many fault types and new ones are found periodically that may not be compatible with \dmaacc's operation. Having few mutant operators could lead to an incomplete assessment of the mutation testing approaches. Incorporating a more diverse set of mutation operators could help correct this, allowing \dmaacc to cover as many fault types as possible.
With only three mutation operators, the study may not be capturing the full range of potential faults and behaviors that can occur in DNNs.
Another threat could be the evaluation metrics, speedup and mutation score error, as they may not fully capture all relevant aspects of mutation testing acceleration.
There could be other important factors, such as memory usage or scalability to very large models, that we did not measure.
The non-deterministic nature of DNN training and some aspects of our approach (e.g. random selection of cluster representatives) could introduce variability in results.
While we attempted to mitigate this through multiple runs, there may still be some effects of randomness.
There is a need for a different method of choosing representatives of mutant clusters rather than one that produces these non-deterministic outputs.
\dmaacc may contain undetected bugs or implementation errors.
These potential issues could affect the accuracy of the mutation analysis process, the clustering algorithms, or the performance measurements. 
Our statistical analysis and conclusions are based on a limited number of models, datasets, mutation operators, and runs.
A larger-scale study with more models and statistical rigor would increase confidence in the generalizability of our findings on the trade-offs between speedup and mutation score error.
Future extensions of this paper will prove beneficial to mitigating these threats.

%% file: sections/related.tex
\section{Related Work}\label{sec:related}
DNN mutation analysis has been applied in many areas~\cite{bib:wang2019adversarial,bib:hu2023muten,bib:hu2019deepmutation++,bib:lin2022robustness,bib:wang2021prioritizing,bib:hu2023aries,bib:sohn2023arachne,bib:wu2022genmunn,bib:ghanbari2023mutation,bib:riccio2021deepmetis,bib:deokuliar2023improving,bib:zohdinasab2024focused,bib:pour2021search,bib:jahangirova2021quality,bib:ghanbari2024decomposition}, and, like its traditional counterpart~\cite{papadakis2019mutation,bib:jia2010analysis}, holds potential for many other applications.
And like its traditional counterpart, DNN mutation analysis is known to be an extremely costly process~\cite{ma2018deepmutation,bib:hu2019deepmutation++,bib:jahangirova2020empirical}.
Motivated by the potentials of DNN mutation analysis, in recent years, several techniques for accelerating DNN mutation analysis has been proposed.

For example, Feng \etal~\cite{bib:feng2022mutation} and Wang \etal~\cite{bib:wang2023fine} took steps to reduce redundant mutants by identifying sufficient subsets of existing mutation operators from the literature~\cite{ma2018deepmutation,bib:shen2018munn}.
Ghanbari \etal~\cite{bib:ghanbari2023mutation} adopts random mutant selection, and the technique presented by Li \etal~\cite{bib:li2022higher} accelerates mutation testing through higher-order mutants akin to higher-order mutation in traditional programs~\cite{bib:jia2009higher}.
The technique presented by Shen \etal~\cite{bib:shen2021boundary}, reduces mutation analysis costs thought test data selection.
It relies on the assumption that mutant of a DNN model are more likely to produce different results for test data points around the decision boundary of the model, so it samples the test data that lie at the decision boundary of the model under test.
Lastly, Incite~\cite{bib:ghanbari2024decomposition} uses neuron clustering to curtail the cost of mutation analysis for the purpose of modular decomposition of DNNs.
In this paper, we studied the effectiveness of this technique and concluded that, while neuron clustering results in greater speedup, clustering at the level of mutants better preserves mutation score.

This paper improves the state of the art by (1) studying the effectiveness of two techniques for reducing the costs of DNN mutation analysis; (2) providing an implementation of the techniques, in a tool named \dmaacc, that can be integrated with existing mutation analysis framekworks~\cite{ma2018deepmutation,bib:humbatova2021deepcrime} with a bit of engineering work.

%% file: sections/confuture.tex
\section{Conclusions}\label{sec:confuture}
This paper presents \dmaacc, a framework for speeding up DNN mutation analysis through neuron and mutant clustering.
\dmaacc implements two methods: (1) neuron clustering to reduce the number of generated mutants and (2) mutant clustering to reduce the number of mutants to be tested by selecting representative mutants for testing.
\dmaacc is publicly available at~\cite{bib:replica}.

This paper further presents an empirical study of two proposed methods using 8 DNN models across 4 popular classification datasets and 2 model architectures.
When compared to vanilla mutation analysis, the results provide empirical evidence that the neuron clustering approach on average speeds up mutation analysis by 69.77\% with an average -26.84\% error in mutation score, while the mutant clustering approach speeds up mutation analysis by 35.31\% with an average 1.96\% error in mutation score.